\documentclass{piner}
\begin{document}

\title{Relativistic Jets in The Radio Reference Frame Image Database I:
Apparent Speeds from the First Five Years of Data}

\author{B.~G.~Piner\altaffilmark{1,3}, M.~Mahmud\altaffilmark{1,4}, A.~L.~Fey\altaffilmark{2}, \& K.~Gospodinova\altaffilmark{1}}

\altaffiltext{1}{Department of Physics and Astronomy, Whittier College,
13406 E. Philadelphia Street, Whittier, CA 90608; gpiner@whittier.edu}

\altaffiltext{2}{U.S. Naval Observatory, 3450 Massachusetts Ave.,
Washington D.C. 20392; afey@usno.navy.mil}

\altaffiltext{3}{Jet Propulsion Laboratory,
California Institute of Technology, 4800 Oak Grove Drive, Pasadena, CA
91109}

\altaffiltext{4}{Department of Physics, University College Cork, Cork, Ireland}

\begin{abstract}
We present the results of an analysis of relativistic jet apparent speeds from VLBI images in the
Radio Reference Frame Image Database (RRFID). 
The RRFID is a database of approximately 4000 images 
of 500 sources compiled from geodetic and astrometric Very Long Baseline Array (VLBA) experiments since 1994.
The images are snapshot VLBI images at 8 and 2~GHz using the VLBA plus up to ten additional
antennas that provide global VLBI coverage. For this paper, we have analyzed the 8~GHz images from the
first five years of the database (1994-1998), for all sources observed at three or more epochs during this time range.
This subset, referred to as the RRFID kinematic survey, comprises 966 images of 87 sources. 
The sources in this subset have an average of 11 epochs of observation
over the years 1994-1998, with the best-observed sources having 19 epochs.
We have measured apparent speeds for a total of 184 jet components in 77 sources, of which
the best-measured 94 component speeds in 54 sources are used in the final analysis. About half of the sources
in the RRFID kinematic survey have not been previously studied with multi-epoch VLBI observations.
A table containing all Gaussian model components that were fit to the observed visibilities 
(a total of 2579 model components) is presented in machine-readable form.

The apparent speed distribution shows a peak at low apparent speeds (consistent with stationary components), a tail
extending out to apparent speeds of about $30c$, and a mean apparent speed of 3.6$c$. 
This apparent speed distribution is statistically consistent with the apparent speed distributions 
found by other radio-selected multi-epoch VLBI surveys.
The fastest measured apparent speeds in the BL Lac objects are slower than in the quasars,
but at a relatively low confidence level of 94\%.
Significant nonradial motion is found for six individual jet components, and 
a tentative detection of accelerated radial motion is reported for three individual components.
There is no significant difference found between the fastest measured apparent speeds in the
EGRET-detected and non-detected sources; however, our samples of those populations are not complete.

A total of 36 of the sources in this paper are also included in the 2~cm VLBA survey by Kellermann et al.,
with similar angular resolution, sensitivity, and time range. For those sources, we present a detailed
component-by-component comparison of the apparent speeds measured by the 2~cm survey and
those measured in this paper. Many of the independent apparent speed measurements agree very well,
but for approximately 25\% of the components we find significant differences in the apparent speeds measured
by the two surveys. The leading cause of these discrepancies
are differences in how the two surveys have identified jet `components' from epoch-to-epoch,
which is influenced by the different epoch spacings in the two surveys.
This is the first such large-scale test of the repeatability of VLBI apparent speed measurements, and it has important
implications for the interpretation of multi-epoch VLBI kinematic results.
\end{abstract}


\section{Introduction}
\label{intro}
Among the most remarkable discoveries made during the early days of the VLBI technique
was the apparent superluminal motion exhibited by the jets of some extragalactic radio sources
(e.g., Whitney et al. 1971; Cohen et al. 1971), which can be explained as a relativistic jet moving nearly along the line of sight 
(e.g., Blandford \& K\"{o}nigl 1979).
Multi-epoch studies of various individual sources began soon after the discovery
of this phenomenon, with different research groups
taking the responsibility for monitoring different sources with ad hoc VLBI arrays.
However, the need was soon recognized for multi-epoch surveys of many sources so that the apparent speeds
of relativistic jets could be studied in a uniform manner, both to learn about the jets themselves, 
and to use in applications ranging from
unified models of AGN to cosmology.
Early efforts to assemble multi-epoch data for many sources relied on collecting single-source results
from the literature (e.g., Vermeulen \& Cohen 1994), but such assemblages can be biased because
they include only the data that observers have elected to publish.
Observing and reducing multiple epochs of VLBI data on many sources
is a time consuming task that became manageable with the advent of dedicated VLBI arrays
such as the National Radio Astronomy Observatory's
Very Long Baseline Array (VLBA)
and the European VLBI Network (EVN),
and the past few years have seen publication of results from two large multi-epoch VLBI surveys
(here `large' is defined as exceeding about 500 images in the survey), which we summarize below.

The Caltech-Jodrell Bank Flat-Spectrum survey (CJF) is a 
complete flux-limited sample of 293 flat-spectrum radio sources,
drawn from the 6~cm and 20~cm Green Bank Surveys.
The CJF survey has obtained 3-5 epochs of VLBI data at 5 GHz
spanning 4-8 years on each of these 293 sources, over the years
1990-2000 (Vermeulen et al. 2003), with a typical separation between
observations of about two years (Britzen et al. 1999).
Some results on the measured apparent speeds, including statistics derived
from 597 component speed measurements in 262 sources, are given by Vermeulen et al. (2003).

The 2~cm survey (Kellermann et al. 1998; Zensus et al. 2002; Kellermann et al. 2004; Kovalev et al. 2005) 
observed a smaller number of sources than the CJF survey,
but at a higher angular resolution because the observations were at a higher frequency of 15 GHz (wavelength of 2~cm).
This survey consisted of multi-epoch observations with the VLBA of over 100 sources between
the years 1994 and 2001. The jet kinematics derived from these observations are presented
by Kellermann et al. (2004) (hereafter K04). That paper presents apparent speeds for 208 jet features
in 110 sources, measured from an average of 6 epochs per source over the years 1994 to 2001,
yielding a typical epoch spacing for each source of about one observation per year.
Since 2001, the 2~cm survey has continued as the MOJAVE survey, with polarization observations
added, and with a somewhat altered source list to ensure a statistically complete sample. 
First epoch results from the MOJAVE survey are presented
by Lister \& Homan (2005) (linear polarization measurements) and Homan \& Lister (2006)
(circular polarization measurements).

In this paper, we present results from a new multi-epoch VLBI kinematic survey, drawn from the U.S. Naval
Observatory's Radio Reference Frame Image Database (RRFID) \footnote 
{The web site for the RRFID is located at http://rorf.usno.navy.mil/RRFID/}.
The Radio Reference Frame Image Database is the result of an ongoing program to 
image radio reference frame sources on a regular basis. 
The goal is to establish a database of images of all radio reference frame 
sources at the same wavelengths as those used for precise astrometry. 
The multi-epoch VLBI data allow the monitoring of sources for variability or structural changes so 
that they can be evaluated for continued suitability as radio reference frame objects. 
RRFID observations are performed with the full ten-station VLBA with the addition of up to ten
geodetic VLBI antennas for global VLBI coverage. Observations are performed 
simultaneously at frequencies of 8 and 2 GHz.
Observations began in 1994 and have continued through the present, however the RRFID image database is
currently well-filled only through the end of 1998, covering the first five years of observations.
At this writing, the database contains 4164 images of 517 sources.
Imaging results from this database have been presented by Fey, Clegg, \& Fomalont (1996);
Fey \& Charlot (1997); and Fey \& Charlot (2000).
The RRFID has also recently begun to include multi-epoch VLBA observations at higher frequencies of
24 and 43~GHz (K- and Q-band)\footnote{http://rorf.usno.navy.mil/RRFID\_KQ/}.

The RRFID is not a flux-limited sample, and membership in the database is limited to those sources
useful for astrometry or geodesy. Sources selected for astrometry and geodesy have historically been
the brightest known compact sources. The source list has evolved over time
as the arrays used evolved and included longer baselines and became
more sensitive, thus rejecting sources previously considered compact
in favor of weaker, more core dominated sources that are known to produce consistent geodetic results.
There has also been an attempt to select sources as
uniformly distributed on the sky as possible --- this can only be
carried so far, but gets easier as the sensitivity to weaker sources increases.
The final result is that while the RRFID contains many observations of well-known sources (for example, BL Lac),
there are other well-known sources (for example, 3C~273) that are not well-represented in the database.
On the other hand, some strong sources that may have been only sparsely observed by the astronomical community
have very good coverage in the RRFID.  Because of this historical evolution of the source list,
the exact nature of the biases that this lack of pre-defined selection criteria may introduce into
statistical quantities calculated from the RRFID is not known.

This paper presents the results of a VLBI kinematic survey (hereafter the RRFID kinematic survey), selected as a subset
of the currently available observations in the RRFID. 
For this paper, we have selected all 8 GHz observations of all
sources observed at 3 or more epochs between the beginning of the
database in 1994 July through the last nearly contiguous epoch in the database in 1998 December ---
this subset covers 19 astrometric VLBA experiments. 
These selection criteria yield a total of 87 sources with a total of 966 8 GHz images.
For these 87 sources, there are then an average of 11 epochs per source over this five year timespan,
with the best observed sources being observed at all 19 epochs.
This survey thus covers a slightly smaller number of sources than the 2~cm survey, but 
with about twice the average number of epochs per source. 
The epoch spacing in the RRFID is not evenly distributed --- there are six astrometric VLBA experiments from
the three years 1994-1996 (yielding an average epoch spacing of about 0.5 years for sources observed at all epochs),
and thirteen astrometric VLBA experiments from the two years 1997-1998
(yielding an average epoch spacing of about 2 months for sources observed at all epochs).
Astrometric VLBA experiments (the RDV series) have continued to use an epoch spacing of about 2 months since 1998.

This survey thus explores the jet kinematics using a much smaller average epoch spacing than
has been used in previous large VLBI surveys, and this is potentially quite important 
(Jorstad et al. (2001) and
Homan et al. (2001) used similarly short time spacing, but for smaller numbers of sources). 
A persistent problem in the interpretation of multi-epoch VLBI images is in the identification
of jet `components' from epoch-to-epoch across the series of images  --- this is arguably the most subjective
step in the usual process of measuring jet apparent speeds. 
A subset of sources may exhibit clear motions
of bright well-separated features, but in other sources there can be ambiguities in the identification
of features that can be influenced by the epoch spacing of the observations.
(In particular, a `strobing' effect can cause a larger number of rapidly moving features to
be interpreted as a smaller number of slower moving features.)
This general problem of component identifications 
is discussed in the context of the CJF survey by Vermeulen et al. (2003).
About half of the sources included in this paper are also included in the 2~cm survey
(many of the other half have not previously been observed with multi-epoch VLBI), and for those sources 
that are common to both surveys we are able to perform
a detailed comparison of our apparent speed measurements with those of K04 over
the same time range, in order to determine if different epoch spacings influence the measurement of jet speeds.
This is the first time known to us that such a large comparison of kinematic results for the same
sources from different VLBI surveys has been attempted, and such a comparison is important for
assessing the repeatability of apparent speed measurements.

We note that the application of geodetic or astrometric VLBI data to astrophysics is not new, see, for example, the
study of geodetic VLBI observations of EGRET blazars by Piner \& Kingham (1997a; 1997b; 1998).
What is different about this application of geodetic VLBI data is that now the data are drawn from observations with the
VLBA that have been designed for accurate imaging of the source structure, so that the images are
of much higher quality than the images in those earlier papers that relied solely on data
from dedicated geodetic antennas.

In this paper we present only the jet apparent speeds from the first five years of 8 GHz data in the RRFID.
Future papers in this series will study other aspects of jet astrophysics from 
both the 8 and 2 GHz data in the RRFID such as:
correlations of the jet apparent speeds with other source properties,
bending of the parsec-scale jets and their misalignment with kiloparsec-scale structures,
transverse structures (or lack thereof) in the parsec-scale jets,
and measurements of jet apparent speeds using an expanded time baseline of 
data (once observations from the years after 1998 are added to the RRFID).

The structure of this paper is as follows: in $\S$\ref{obs} we describe the RRFID observations in detail,
in $\S$\ref{data} we describe the procedures used in calibrating, imaging, and model fitting the data,
in $\S$\ref{results} we present the kinematic results, and in $\S$\ref{comp} we compare those kinematic
results to those obtained by the 2~cm survey (K04) for the common sources.
Throughout this paper we assume cosmological parameters 
$H_{0}=71$ km s$^{-1}$ Mpc$^{-1}$, $\Omega_{m}=0.27$, and $\Omega_{\Lambda}=0.73$.
When results from other papers are quoted,
they have been converted to this cosmology.

\section{Observations}
\label{obs}
Observations were made using the 10 antennas of the VLBA (Napier et~al. 1994)
of the National Radio Astronomy Observatory (NRAO)
\footnote{The
National Radio Astronomy Observatory (NRAO) is operated by Associated
Universities, Inc., under cooperative agreement with the National
Science Foundation.}, 
along with an array consisting of up to 7 geodetic
antennas (GILCREEK -- Fairbanks, AK; NRAO20 -- Green Bank, WV; 
KOKEE -- Kokee Park, HI; MEDICINA -- Medicina, Italy;
NYALES20 -- Ny Alesund, Norway; ONSALA60 -- Onsala, Sweden; \&
WESTFORD -- Westford, MA).
Eight intermediate frequencies (IFs) were recorded
simultaneously, each eight MHz wide, with four at S-band (2.24, 2.27,
2.36, \& 2.38 GHz) and four at X-band (8.41, 8.48, 8.79, \& 8.90 GHz)
for a total bandwidth of 32~MHz in each frequency band.  Observations
were made in a dual-frequency bandwidth synthesis mode to facilitate
delay measurements for geodesy and astrometry. The multiplicity of
channels allows for the determination of a precise group delay (Rogers
1970), while simultaneous observations in two bands allows for an
accurate calibration of the frequency dependent propagation delay
introduced by the ionosphere. Results of the precise geodesy and
astrometry afforded by these observations has been presented elsewhere
(e.g., Petrov \& Ma 2003; Fey et~al. 2004). Observations in this mode
also allow simultaneous dual-frequency imaging, which is the focus of
the work discussed here.
Of order 100 sources are observed in a single 24-hour experiment, for an
average time on source per experiment of about 15 minutes.
This time on source is divided into scans of a minute to a few minutes in length
that are spread throughout the 24-hour observing period.

Table~\ref{obstab} shows the 19 VLBA experiments that are included in this paper.
As can be seen from this table, the earlier experiments (1994-1996) are spaced
more sporadically in time and used only the 10-station VLBA. The later VLBA experiments,
corresponding with the beginning of the RDV series in 1997, are spaced
roughly every two months in time, and these used the full VLBA plus up to seven
geodetic antennas. The RDV experiment series has continued to observe every two
months through the present (and is currently up to RDV57), but the epochs
after RDV12 are not yet fully integrated into the RRFID.

\begin{table*}[!t]
\caption{Observation Log}
\vspace{-0.20in}
\label{obstab}
\begin{center}
\begin{tabular}{l c l c} \colrule \colrule
& VLBA Observation & & Image \\
Epoch & Code & Antennas$^{a}$ & Reference \\ \colrule  
1994JUL08 & BR005  & VLBA                & 1 \\
1995APR12 & BR025  & VLBA                & 2 \\
1995JUL24 & RDGEO2 & VLBA                & 3 \\
1995OCT02 & RDGEO3 & VLBA                & 3 \\
1995OCT12 & BF012  & VLBA                & 2 \\
1996APR23 & BE010a & VLBA                & 3 \\
1997JAN10 & BF025a & VLBA                & 4 \\
1997JAN11 & BF025b & VLBA                & 4 \\
1997JAN30 & RDV01  & VLBA+GcGnKkMcOnWf   & 3 \\
1997MAR31 & RDV02  & VLBA+GcGnKkMcOnWf   & 3 \\
1997MAY19 & RDV03  & VLBA+GcGnKkMcOnWf   & 3 \\
1997JUL24 & RDV04  & VLBA+GcGnKkMcOnWf   & 3 \\
1997SEP08 & RDV05  & VLBA+GcGnKkOnWf     & 3 \\
1997DEC17 & RDV06  & VLBA+GcGnKkMcOnWf   & 3 \\
1998FEB09 & RDV07  & VLBA+GcGnKkMcNyOnWf & 3 \\
1998APR15 & RDV08  & VLBA+GcGnKkMcNyOnWf & 3 \\
1998JUN24 & RDV09  & VLBA+GcGnKkMcNyOnWf & 3 \\
1998AUG10 & RDV10  & VLBA+GcGnKkMcNyOn   & 3 \\
1998DEC21 & RDV12$^{b}$  & VLBA+GcGnKkMcNyWf   & 3 \\ \colrule  
\end{tabular}
\end{center}
$a$: Gc - GILCREEK; Fairbanks, AK USA,
Gn - NRAO20; Green Bank, WV USA,
Kk - KOKEE; Kokee Park, HI USA,
Mc - MEDICINA; Medicina, Italy,
Ny - NYALES20; Ny Alesund, Norway,
On - ONSALA60; Onsala, Sweden,
Wf - WESTFORD; Westford, MA USA\\
$b$: RDV12 is not technically contiguous with RDV10, but since only the single
experiment RDV11 is missing in between,
we chose to include RDV12 in this paper. At the time of this writing the
next experiment in the RRFID after RDV12 was RDV31 from 2002 January 16.
\\ [5pt]
References. --- (1) Fey et al. (1996);
(2) Fey \& Charlot (1997);
(3) http://rorf.usno.navy.mil/RRFID/;
(4) Fey \& Charlot (2000).
\end{table*}

The sample selection for this paper was chosen as the set of all sources that were
observed at 3 or more epochs in the VLBA experiments listed in Table~\ref{obstab}.
Since this paper is primarily concerned with measuring the jet apparent speeds, we
considered only the 8 GHz images in this paper, saving the lower-resolution 2 GHz observations
for future papers. These selection criteria yielded a sample of 87 sources with a total
of 966 8~GHz images. The best observed sources were observed at all 19 epochs, and the
average number of epochs per source is 11.
The list of the 87 sources in the current RRFID kinematic survey is given in Table~\ref{sources}. 
This table gives the name of the source in IAU format, other common names, the number of VLBI
epochs in the RRFID kinematic survey, the redshift if known, and the source optical type (BL Lac object,
quasar, or galaxy) from the V\'{e}ron-Cetty \& V\'{e}ron (2003) catalog.

\begin{table*}[!t]
\caption{Sources in the RRFID Kinematic Survey}
\vspace{-0.20in}
\label{sources}
\begin{center}
\begin{tabular}{l l c c c | l l c c c} \colrule \colrule  
& \multicolumn{1}{c}{Common} & Number of & Optical & & & \multicolumn{1}{c}{Common} & Number of & Optical & \\ 
\multicolumn{1}{c}{Source} & \multicolumn{1}{c}{Name} & Epochs & Class$^{a}$ & $z$ &
\multicolumn{1}{c}{Source} & \multicolumn{1}{c}{Name} & Epochs & Class$^{a}$ & $z$ \\ \colrule  
0003-066 &         & 12 & B      & 0.35  & 1128+385 &         & 15 & Q      & 1.73 \\
0014+813 &         & 12 & Q      & 3.37  & 1144-379 &         & 10 & Q(HP)  & 1.05 \\
0048-097 &         & 15 & B      & ...   & 1145-071 &         & 12 & Q      & 1.34 \\
0059+581 &         & 14 & Q$^{b}$& 0.64  & 1156+295 &         & 12 & Q(HP)  & 0.73 \\
0104-408 &         & 10 & Q      & 0.58  & 1219+044 &         & 11 & Q      & 0.97 \\
0111+021 &         &  8 & B      & 0.05  & 1228+126 & M87     & 12 & G      & 0.004 \\
0119+041 &         & 14 & Q(HP)  & 0.64  & 1253-055 & 3C~279  & 3  & Q(HP)  & 0.54 \\
0119+115 &         & 12 & Q(HP)  & 0.57  & 1255-316 &         & 5  & Q      & 1.92 \\
0133+476 &         & 13 & Q(HP)  & 0.86  & 1300+580 &         &  9 & U      & ...  \\
0146+056 &         &  4 & Q      & 2.35  & 1308+326 &         & 15 & Q(HP)  & 1.00 \\
0201+113 &         & 13 & Q      & 3.61  & 1313-333 &         & 12 & Q      & 1.21 \\
0202+149 &         & 13 & G      & 0.41  & 1334-127 &         & 13 & Q(HP)  & 0.54 \\
0229+131 &         & 15 & Q      & 2.06  & 1351-018 &         &  7 & Q      & 3.71 \\
0234+285 &         & 13 & Q(HP)  & 1.21  & 1357+769 &         & 15 & Q$^{b}$& ...  \\
0238-084 & NGC1052 & 10 & G      & 0.005 & 1404+286 & OQ~208  & 11 & G      & 0.08 \\
0336-019 & CTA~26  & 13 & Q(HP)  & 0.85  & 1418+546 &         &  8 & B      & 0.15 \\
0402-362 &         & 10 & Q      & 1.42  & 1424-418 &         &  8 & Q(HP)  & 1.52 \\
0430+052 & 3C~120  & 11 & G      & 0.03  & 1451-375 &         &  9 & Q      & 0.31 \\
0454-234 &         & 14 & Q(HP)  & 1.00  & 1514-241 &         & 11 & B      & 0.05 \\
0458-020 &         & 12 & Q(HP)  & 2.29  & 1606+106 &         & 14 & Q      & 1.23 \\
0528+134 &         & 14 & Q      & 2.06  & 1611+343 &         & 13 & Q      & 1.40 \\
0537-441 &         & 10 & Q(HP)  & 0.89  & 1622-253 &         & 13 & Q      & 0.79 \\
0552+398 &         & 18 & Q      & 2.37  & 1638+398 & NRAO512 & 14 & Q(HP)  & 1.66 \\
0556+238 &         &  8 & U      & ...   & 1652+398 & Mrk 501 &  3 & B      & 0.03 \\
0642+449 &         & 12 & Q      & 3.40  & 1726+455 &         & 10 & Q      & 0.72 \\
0718+793 &         &  4 & U      & ...   & 1739+522 &         & 14 & Q(HP)  & 1.38 \\
0727-115 &         & 19 & Q      & 1.59  & 1741-038 &         & 15 & Q(HP)  & 1.05 \\
0742+103 &         & 9  & G$^{b}$& 2.62  & 1745+624 &         & 12 & Q      & 3.89 \\
0749+540 &         & 3  & B      & ...   & 1749+096 &         & 19 & Q(HP)  & 0.32 \\
0804+499 &         & 13 & Q(HP)  & 1.43  & 1803+784 &         & 12 & Q(HP)  & 0.68 \\
0805+410 &         & 3  & Q      & 1.42  & 1908-201 &         & 11 & Q      & 1.12 \\
0823+033 &         & 14 & B      & 0.51  & 1921-293 &         & 14 & Q(HP)  & 0.35 \\
0851+202 & OJ~287  & 15 & B      & 0.31  & 1928+738 &         & 3  & Q      & 0.30 \\
0919-260 &         & 12 & Q      & 2.30  & 1954-388 &         & 11 & Q(HP)  & 0.63 \\
0920-397 &         & 10 & Q      & 0.59  & 1958-179 &         & 3  & Q(HP)  & 0.65 \\
0923+392 & 4C39.25 & 14 & Q      & 0.70  & 2052-474 &         & 4  & Q      & 1.49 \\
0953+254 & OK~290  & 8  & Q      & 0.71  & 2136+141 &         & 8  & Q      & 2.43 \\
0955+476 &         & 14 & Q      & 1.87  & 2145+067 &         & 19 & Q      & 0.99 \\
1004+141 & 	   & 9  & Q      & 2.71  & 2200+420 & BL LAC  & 12 & B      & 0.07 \\
1022+194 &         & 3  & Q      & 0.83  & 2230+114 & CTA~102 & 6  & Q(HP)  & 1.04 \\
1034-293 &         & 13 & Q(HP)  & 0.31  & 2234+282 &         & 14 & Q(HP)  & 0.80 \\
1044+719 &         & 14 & Q      & 1.15  & 2243-123 &         & 12 & Q(HP)  & 0.63 \\
1101+384 & Mrk 421 & 12 & B      & 0.03  & 2255-282 &         & 11 & Q      & 0.93 \\
1124-186 &         & 12 & Q      & 1.05 \\ \colrule  
\end{tabular}
\end{center}
$a$: Optical class from V\'{e}ron-Cetty \& V\'{e}ron (2003).
Q=quasar, B=BL Lac object, G=galaxy, HP=high polarization, U=unidentified.\\
$b$: ID from NED. (Source not in the V\'{e}ron-Cetty \& V\'{e}ron (2003) catalog.)
\end{table*}

\section{Data Analysis}
\label{data}
\subsection{AIPS Calibration}
The raw data bits were correlated with the VLBA correlator at the
Array Operations Center in Socorro, New Mexico.  The correlated data
were calibrated and corrected for residual delay and delay rate using
the NRAO Astronomical Image Processing System (AIPS).  Initial
amplitude calibration for each of the 8 IFs was accomplished using
system temperature measurements taken during the observations combined
with station supplied gain curves.  Fringe-fitting was done in
AIPS using solution intervals equal to the scan durations and a point
source model in all cases.  After correction for residual delay and
delay rate, the data were written to FITS disk files.  All subsequent
processing was carried out using the Caltech VLBI imaging software,
primarily DIFMAP.

\subsection{Imaging}
The visibility data for each frequency band were self-calibrated,
Fourier inverted, and CLEANed using DIFMAP in an automatic mode
(Shepherd, Pearson, \& Taylor 1995). DIFMAP combines the visibilities
for each IF of an observation in the $(u,v)$-plane during gridding, taking
into account frequency differences. However, DIFMAP makes no attempt
to correct for spectral index effects. The spanned bandwidths of the
four IFs in each band are 0.1 GHz (6\% fractional bandwidth) at S-band
and 0.5 GHz (6\% fractional bandwidth) at X-band, so it is possible that spectral
index changes in the core or jet components could cause small errors in the
fitted component positions, because the model components are assumed to have zero
spectral index. We investigated the magnitude of this effect by using the
AIPS task UVMOD to generate components of known position and spectral index,
sampled with a typical RRFID $(u,v)$-plane coverage. The typical error introduced
for spectral index changes of about 1 in the core is only a few $\mu$as, confirming
that spectral index effects on position measurements are negligible for our
relatively small 6\% bandspead.

After phase self-calibration with a point source model, the 4 second
correlator records were coherently averaged to 12 second records and
then edited.
Amplitude calibration at each frequency band was improved through
observations of a strong, compact source.  A single amplitude gain
correction factor was derived for each antenna for each IF, based on
fitting a simple Gaussian source model to the visibility data of these
compact sources after applying only the initial calibration based on
the measured system temperatures and gain curves.  Gain correction
factors were calculated based on the differences between the observed
and model visibilities.  The resulting set of amplitude gain
correction factors was then applied to the visibility of all
sources. The absolute flux density scale of the data has not been
investigated but is estimated to be within 10-20\%.

The data were self-calibrated following the hybrid-mapping technique
(Pearson \& Readhead 1984) to correct for residual amplitude and phase
errors. The data were initially phase self-calibrated and mapped using
uniform weighting in the u,v-plane before switching to natural
weighting after several iterations.  A point source model was used as
a starting model for the iterative procedure in all cases.
Convergence was defined basically as the iteration when the peak in
the residual image became less than a specified factor times the
root-mean-square (rms) noise of the residual image from the previous
iteration. Sources with emission structure too complex or too extended
for the automatic imaging script to handle were imaged by hand,
i.e. in an interactive mode, following the same prescription as that
for the automatic mode. Convergence for these sources was subjective
and was based on the iteration at which it was judged that further
self-calibration would not significantly improve the resultant image.

Far too many VLBA images (966 in total) were used for this paper to present
them all in printed form here. A subset of the RRFID images used in this paper has been presented in
printed form by Fey et al. (1996), Fey \& Charlot (1997), and Fey \& Charlot (2000).
In addition, the final CLEAN images are all publicly available from the Radio Reference Frame
Image Database (RRFID) at http://rorf.usno.navy.mil/RRFID/.
In Figure~1, we show a sample of three $(u,v)$-plane coverages along with their corresponding images,
in order to show typical $(u,v)$-plane coverages obtained for high, medium, and low declination
sources, respectively.

\begin{figure*}
\begin{center}
\includegraphics[scale=0.75]{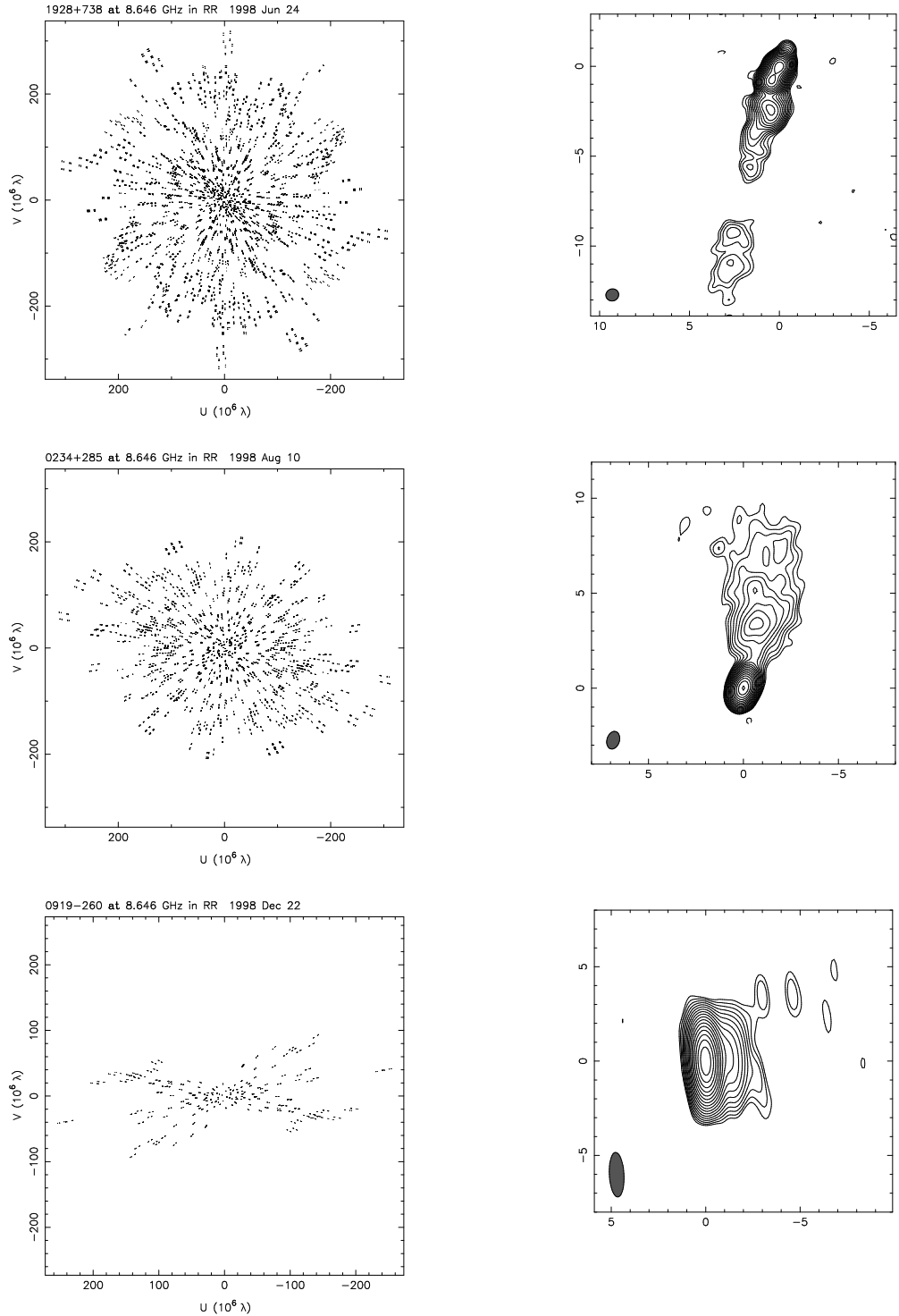}
\end{center}
\caption{Sample $(u,v)$-plane coverages and associated images for a high-declination (1928+738),
medium-declination (0234+285), and low-declination (0919-260) source from the RRFID.
Image axes are in milliarcseconds. 
For 1928+738 the image parameters are: lowest contour 4.3 mJy beam$^{-1}$,
peak flux 1.07 Jy beam$^{-1}$, and beam size 0.72 by 0.67 mas in position angle $-68\arcdeg$.
For 0234+285 the image parameters are: lowest contour 2.8 mJy beam$^{-1}$,
peak flux 1.04 Jy beam$^{-1}$, and beam size 0.95 by 0.66 mas in position angle $-17\arcdeg$.
For 1928+738 the image parameters are: lowest contour 3.7 mJy beam$^{-1}$,
peak flux 1.25 Jy beam$^{-1}$, and beam size 2.36 by 0.78 mas in position angle $3\arcdeg$.
Lowest contours are set to three times the rms noise level in the residual maps.
All other contours are factors of $\sqrt{2}$ higher than the previous contour.}
\end{figure*}

Because a major section of this paper is a comparison of our jet apparent speed measurements
with the 2~cm survey speed measurements for the sources in common, we compare
here the factors that influence the image dynamic range and angular resolution in 
these two surveys.
The VLBA beam size is about a factor of two larger at 8 GHz than at 15 GHz, thus the RRFID
images that are VLBA-only (the first eight epochs in Table~\ref{obstab}) have about a factor
of two worse angular resolution than the 2~cm survey images. However, the global VLBI beam size at
8 GHz is approximately equal to the VLBA beam size at 15 GHz, so that the later epochs in the RRFID
that used global VLBI networks have similar angular resolution to the 2~cm survey observations.
The RRFID observations use about one-quarter of the time on source and half the bandwidth
of the 2~cm survey observations, and this implies a loss of sensitivity by a factor of a few relative
to the 2~cm survey. However, these factors are compensated for by the greater number of antennas
in most of the RRFID observations, the lower SEFD (System Equivalent Flux Density) of the antennas at 8 GHz
relative to 15 GHz, and the higher flux of the jet components at 8 GHz.
The result of all of these factors is that the dynamic range of the jet detections
in the two surveys is similar.
Because of their very different setups, it is fortuitous that the RRFID images and the 2~cm survey images
are similar in angular resolution and jet dynamic range; this similarity greatly facilitates the comparisons
between the results of the two surveys in $\S$\ref{comp}. 

\subsection{Model Fitting}
\label{modelfitting}
Gaussian models were fit to the self-calibrated {\it visibility
data} on a source-by-source basis using DIFMAP in an interactive
mode. The number of Gaussian components and the choice between
elliptical components or circular components was subjective, but
motivated by consideration of the simplicity of the resulting model.
Thus, elliptical components were used sparingly, and only to represent the core
or a very bright jet component when the residuals remaining from a circular Gaussian
fit were so large as to hinder further model fitting using the residual map.
Although the agreement
between the fitted models and the data is not as good as that produced
by the hybrid images (models with many CLEAN components), inspection
of plots of residuals in the image plane, after subtracting the
Gaussian models from the visibility data, revealed that the Gaussian
models generally describe the visibility data quite well.  However,
because of incomplete sampling in the $(u,v)$-plane, these models may not
be unique.  They represent only one {\it possible} deconvolution of
complex source structure. Such deconvolutions can be misleading.

The models were fit to the self-calibrated visibility data corresponding
to the publicly available images in the Radio Reference Frame Image Database, no
further processing of the visibility data was performed, so others should be able
to reproduce the models given here from the publicly available data. 
The models published here were done independently from the previously published models 
of five epochs of RRFID data by Fey et al. (1996), 
Fey \& Charlot (1997), and Fey \& Charlot (2000),
because it was desired that all model fits be done in a uniform manner for this paper. 
A comparison of the model fits for those five epochs that were previously fit by Fey et al. (1996),
Fey \& Charlot (1997), and Fey \& Charlot (2000) shows that in the vast majority ($\sim90\%$) of cases there is agreement
in the fitted positions of common components within the errors. Any difference between these model fits 
lies in the presence of an additional component or components in one or the other of the fits, i.e, 
any difference results from a differing decision of when to stop adding components to the residual map;
and in these cases one model is basically a subset of the other. In a small number ($\sim10\%$) of cases the corresponding
model fits give significantly different component positions; these are cases where multiple deconvolutions of the source
structure are possible, and such cases are discussed further in the context of comparisons with the 2~cm survey
results in $\S$~\ref{comp}.

The model fits for the entire RRFID kinematic survey to date are presented in machine-readable 
form in Table~\ref{mfit}, which contains a total 2579 Gaussian components.
This machine-readable Table~\ref{mfit} should be suitable for script-based processing 
by other investigators.
Some sources listed in Table~\ref{sources} do not have model fits given in
Table~\ref{mfit}, for the following reasons.
\begin{enumerate}
\item{Lack of $z$: six sources (0048-097, 0556+238, 0718+793, 0749+540, 1300+580, 1357+769)
did not have a measured redshift at the time of this writing. Because no apparent speed
could be determined for these sources, we did not model them.}
\item{Complexity: three sources (0238-084, 0430+052, 1404+286) had 8 GHz structures that were two-sided 
(hindering identification of the core), and/or
so smooth and complex at 8 GHz that we were not able to reliably follow components from epoch to epoch.}
\item{Lack of a jet: one source (2052-474) was modeled as only a core component at all epochs.}
\end{enumerate}
In all, 77 of the 87 sources listed in Table~\ref{sources} have model fits in Table~\ref{mfit}.

\begin{table*}
\caption{Gaussian Models}
\vspace{-0.20in}
\label{mfit}
\begin{center}
\begin{tabular}{l r r r r r r c c c r r r} \colrule   \colrule  
& \multicolumn{1}{c}{$S$\tablenotemark{a}} & \multicolumn{1}{c}{$r$\tablenotemark{b}} &
\multicolumn{1}{c}{PA\tablenotemark{b}} &
\multicolumn{1}{c}{$a$\tablenotemark{c}} &
& \multicolumn{1}{c}{$\phi$\tablenotemark{c}} & & & & \multicolumn{1}{c}{$a_{beam}$\tablenotemark{f}}
& \multicolumn{1}{c}{$b_{beam}$\tablenotemark{f}} & \multicolumn{1}{c}{$\theta_{beam}$\tablenotemark{f}} \\
\multicolumn{1}{c}{Source}  
& \multicolumn{1}{c}{(Jy)} & \multicolumn{1}{c}{(mas)} &
\multicolumn{1}{c}{(deg)}
& \multicolumn{1}{c}{(mas)} & \multicolumn{1}{c}{$(b/a)$\tablenotemark{c}} & \multicolumn{1}{c}{(deg)}
& Type\tablenotemark{d} & Epoch & \multicolumn{1}{c}{Comp.\tablenotemark{e}} & 
\multicolumn{1}{c}{(mas)} & \multicolumn{1}{c}{(mas)} & \multicolumn{1}{c}{(deg)} \\ \colrule  
0003-066 & 1.599 & 0.079 & 148.3   & 0.633 & 0.387 & $-$16.3 & 1 & 1995.78 & 0 & 2.29 & 0.95 & $-$1.1 \\
0003-066 & 0.645 & 1.040 & $-$60.5 & 1.384 & 1.000 &     0.0 & 1 & 1995.78 & 3 & 2.29 & 0.95 & $-$1.1 \\
0003-066 & 0.156 & 5.145 & $-$74.5 & 3.222 & 1.000 &     0.0 & 1 & 1995.78 & 1 & 2.29 & 0.95 & $-$1.1 \\
0003-066 & 1.209 & 0.032 & 114.2   & 0.529 & 0.000 &    21.2 & 1 & 1997.08 & 0 & 2.03 & 0.75 & $-$5.8 \\
0003-066 & 0.225 & 0.786 & $-$48.9 & 0.520 & 1.000 &     0.0 & 1 & 1997.08 & 3 & 2.03 & 0.75 & $-$5.8 \\
0003-066 & 0.194 & 2.131 & $-$71.1 & 1.416 & 1.000 &     0.0 & 1 & 1997.08 & 2 & 2.03 & 0.75 & $-$5.8 \\
0003-066 & 0.083 & 5.586 & $-$75.2 & 2.455 & 1.000 &     0.0 & 1 & 1997.08 & 1 & 2.03 & 0.75 & $-$5.8 \\ \colrule  
\end{tabular}
\end{center}
NOTE. -- Table~\ref{mfit} is published in its entirety in machine-readable form
in the electronic edition of the journal.
A portion is shown here for guidance regarding
its form and content.\\
$a$: Flux density of the component.\\
$b$: $r$ and PA are the polar coordinates of the
Gaussian center.
PA is measured from north through east.\\
$c$: $a$ and $b$ are the FWHM of the major and minor axes of the Gaussian,
and
$\phi$ is the position angle of the major axis.\\
$d$: Component type for the DIFMAP `modelfit' command. Type 1 indicates a Gaussian component.\\
$e$: Component `0' indicates the presumed core. Other components are numbered from 1 to 6,
from the outermost component inward. A component ID of `99' indicates a flagged component not used in the analysis.\\
$f$: $a_{beam}$, $b_{beam}$, and $\theta_{beam}$ are the major axis FWHM, 
minor axis FWHM, and position angle of the major axis of the
naturally weighted restoring beam.\\
\end{table*}

The second through the eighth columns of Table~\ref{mfit} correspond directly to the DIFMAP modelfit results,
and are suitable for reading directly into DIFMAP with the `rmodel' command.
Positions in Table~\ref{mfit} have {\em not} been shifted to place the core at the origin, so that the positions
in Table~\ref{mfit} will correspond directly to positions on the publicly available RRFID images.
The tenth column of Table~\ref{mfit} contains the component identification,
the core is identified as component 0, and other jet components are identified as component
1 up to component 6, from the outermost component inward.
We identify the core in each source as
the compact component at the end of the one-sided jet structure --- often,
but not always, it is also the brightest component.
As noted above, we excluded any sources known to show two-sided VLBA structures at these scales.
Identifications of other components from epoch to epoch were done through continuity in radial position,
flux, and position angle; this was facilitated by the dense time coverage of the RRFID during 1997 and 1998.
However, we note that often the identification scheme adopted represents what we considered to be the
most likely of several possible scenarios, and that different identification scenarios can lead to
different measured jet properties.
In cases where a model component could not be directly identified with model components seen at other
epochs (about 5\% of the total fitted components), it was given an identification of `99' in Table~\ref{mfit}
to flag it as a model component not used in the analysis.
This typically happened when a somewhat lower resolution image blended together what was seen as two
separate components in other model fits, or when an
extended low-dynamic-range component was detected in only a few images with a poorly constrained position.

All components were modeled as Gaussians (the only extended brightness distribution fully supported in DIFMAP), 
and it is worth considering whether assuming a different
brightness distribution would have any effect on the measured apparent speeds.
We investigated this by modeling a test source (chosen to be 1308+326 because of the very small scatter
of the measured Gaussian component positions about the best-fit line) a second time with an alternate brightness
distribution --- the core was modeled as a uniformly bright disk and the single jet component as an
optically thin sphere. The measured apparent speed differed by about 5\% between the two cases, but considering
the errors on the fit, this difference was not significant.
Where the assumed brightness distribution will have a large effect is on the measured sizes
of the components (Pearson 1995), but the measured sizes are not used in this paper.

\section {Results}
\label{results}
\subsection{Apparent Speeds}
\label{appspeed}
Apparent radial proper motions for the jet components in Table~\ref{mfit} were derived
from linear least-squares fits to the separation of the jet components from the VLBI core
versus time, for components that were observed at three or more epochs.
The proper motions were then converted to apparent radial speeds using the cosmology given in $\S$\ref{intro}.
These fits are shown in Figure~2, and the results are tabulated in Table~4.
A total of 184 component motions in 77 sources are tabulated in Table~4. 
Note that the disappearance of a component on Figure~2 followed by its re-appearance at later epochs
does not imply the literal disappearance and re-appearance of the
component on the images.  Rather, the component is usually present at a marginally significant level
in the intervening images, but is not significant enough to have its properties well constrained by the
model fitting. 

\begin{table*}
\caption{Apparent Component Speeds}
\vspace{-0.20in}
\begin{center}
\begin{tabular}{l c r r c c c c} \colrule \colrule  
& & \multicolumn{1}{c}{$\mu$} & & & Distinct\tablenotemark{b} & Quality\tablenotemark{c} & 2~cm\tablenotemark{d} \\
\multicolumn{1}{c}{Source} & Comp. &
\multicolumn{1}{c}{(mas yr$^{-1}$)} & \multicolumn{1}{c}{$\beta_{app}$} &
$\xi$\tablenotemark{a}  & Component & Code & Survey \\ \colrule  
0003-066 & 1 & $0.308\pm0.091$  & $6.6\pm2.0$    & 2 & Y & G & Y \\
         & 2 & $0.524\pm0.124$  & $11.3\pm2.7$   & 2 & Y & G &   \\
         & 3 & $-0.020\pm0.029$ & $-0.4\pm0.6$   & 4 & Y & E &   \\
0014+813 & 1 & $-0.083\pm0.050$ & $-8.9\pm5.4$   & 3 & N & G & N \\
0059+581 & 1 & $-0.071\pm0.067$ & $-2.6\pm2.5$   & 2 & Y & E & N \\
         & 2 & $0.106\pm0.034$  & $4.0\pm1.2$    & 3 & N & G &   \\
         & 3 & $-0.002\pm0.022$ & $-0.1\pm0.8$   & 4 & N & G &   \\
         & 4 & $0.025\pm0.103$  & $0.9\pm3.8$    & 3 & N & F &   \\
0104-408 & 1 & $-0.694\pm0.883$ & $-23.7\pm30.2$ & 1 & N & P & N \\
0111+021 & 1 & $0.006\pm0.238$  & $0.0\pm0.7$    & 2 & Y & P & N \\
	 & 2 & $-0.074\pm0.056$ & $-0.2\pm0.2$   & 4 & Y & E &   \\
	 & 3 & $0.008\pm0.029$  & $0.0\pm0.1$    & 5 & N & G &   \\
0119+041 & 1 & $0.013\pm0.018$  & $0.5\pm0.7$    & 4 & N & G & Y \\
0119+115 & 1 & $0.050\pm0.049$  & $1.7\pm1.7$    & 4 & N & G & N \\
         & 2 & $-0.018\pm0.075$ & $-0.6\pm2.5$   & 3 & N & G &   \\
0133+476 & 1 & $-0.052\pm0.093$ & $-2.5\pm4.4$   & 2 & Y & G & Y \\
         & 2 & $0.366\pm0.047$  & $17.2\pm2.2$   & 4 & N & G &   \\
         & 3 & $0.385\pm0.081$  & $18.0\pm3.8$   & 4 & N & F &   \\
0146+056 & 1 & $0.068\pm0.098$  & $6.2\pm8.9$    & 3 & N & F & N \\
0201+113 & 1 & $0.103\pm0.055$  & $11.4\pm6.1$   & 3 & N & G & N \\
         & 2 & $0.047\pm0.037$  & $5.2\pm4.2$    & 4 & N & G &   \\
0202+149 & 1 & $-0.053\pm0.034$ & $-1.3\pm0.8$   & 4 & Y & E & Y \\
         & 2 & $0.062\pm0.033$  & $1.5\pm0.8$    & 4 & N & G &   \\
0229+131 & 1 & $-0.202\pm0.135$ & $-17.1\pm11.4$ & 3 & N & F & N \\ 
         & 2 & $0.156\pm0.090$  & $13.2\pm7.6$   & 2 & Y & G &   \\ 
         & 3 & $0.384\pm0.107$  & $32.5\pm9.1$   & 2 & N & F &   \\ 
         & 4 & $-0.066\pm0.067$ & $-5.6\pm5.7$   & 4 & N & G &   \\
0234+285 & 1 & $0.137\pm0.091$  & $8.3\pm5.5$    & 2 & Y & G & Y \\
         & 2 & $0.159\pm0.023$  & $9.6\pm1.4$    & 4 & Y & E &   \\
         & 3 & $0.068\pm0.103$  & $4.1\pm6.2$    & 2 & N & F &   \\
         & 4 & $0.013\pm0.388$  & $0.8\pm23.5$   & 3 & N & P &   \\
0336-019 & 1 & $0.436\pm0.219$  & $20.3\pm10.2$  & 2 & Y & P & Y \\
         & 2 & $0.151\pm0.093$  & $7.0\pm4.4$    & 2 & N & F &   \\
         & 3 & $0.183\pm0.037$  & $8.5\pm1.7$    & 3 & Y & E &   \\
0402-362 & 1 & $0.386\pm0.152$  & $26.0\pm10.2$  & 3 & Y & P & N \\
         & 2 & $0.032\pm0.083$  & $2.2\pm5.6$    & 4 & N & F &   \\
0454-234 & 1 & $0.116\pm0.130$  & $6.2\pm6.9$    & 2 & N & F & N \\ 
0458-020 & 1 & $-0.558\pm0.232$ & $-50.1\pm20.8$ & 2 & Y & P & Y \\
         & 2 & $-0.019\pm0.044$ & $-1.7\pm3.9$   & 4 & Y & E &   \\
0528+134 & 1 & $-0.166\pm0.111$ & $-14.1\pm9.4$  & 2 & Y & G & Y \\ 
         & 2 & $-0.034\pm0.050$ & $-2.9\pm4.2$   & 3 & N & G &   \\ 
         & 3 & $0.016\pm0.023$  & $1.3\pm1.9$    & 4 & N & G &   \\ 
0537-441 & 1 & $1.109\pm0.478$  & $53.7\pm23.1$  & 1 & N & P & N \\
0552+398 & 1 & $-0.003\pm0.006$ & $-0.2\pm0.6$   & 5 & N & G & N \\ 
0642+449 & 1 & $0.172\pm0.112$  & $18.6\pm12.2$  & 2 & Y & G & Y \\
         & 2 & $0.037\pm0.014$  & $4.0\pm1.5$    & 5 & N & G &   \\
0727-115 & 1 & $-0.035\pm0.044$ & $-2.5\pm3.2$   & 3 & Y & E & Y \\
         & 2 & $0.054\pm0.065$  & $3.9\pm4.7$    & 3 & N & G &   \\
         & 3 & $0.011\pm0.062$  & $0.8\pm4.5$    & 5 & N & F &   \\
0742+103 & 1 & $0.078\pm0.115$  & $7.5\pm11.1$   & 3 & N & F & Y \\
         & 2 & $-0.071\pm0.078$ & $-6.8\pm7.5$   & 3 & N & G &   \\
         & 3 & $-0.037\pm0.018$ & $-3.5\pm1.8$   & 5 & N & G &   \\
0804+499 & 1 & $0.195\pm0.080$  & $13.2\pm5.4$   & 2 & Y & E & Y \\ 
         & 2 & $-0.228\pm0.257$ & $-15.5\pm17.4$ & 2 & N & P &   \\ 
0805+410 & 3 & $-0.073\pm0.322$ & $-4.9\pm21.7$  & 2 & N & P & N \\
0823+033 & 1 & $-0.016\pm0.056$ & $-0.5\pm1.7$   & 3 & Y & E & Y \\
         & 2 & $0.019\pm0.027$  & $0.6\pm0.8$    & 4 & N & G &   \\ 
0851+202 & 1 & $-0.039\pm0.109$ & $-0.7\pm2.1$   & 1 & N & F & Y \\ 
         & 2 & $0.296\pm0.082$  & $5.7\pm1.6$    & 4 & N & P &   \\ 
         & 3 & $0.386\pm0.043$  & $7.4\pm0.8$    & 3 & N & G &   \\ 
         & 4 & $-0.031\pm0.059$ & $-0.6\pm1.1$   & 4 & N & G &   \\ 
0919-260 & 1 & $0.565\pm0.125$  & $50.8\pm11.3$  & 2 & N & F & N \\
         & 2 & $0.038\pm0.036$  & $3.4\pm3.2$    & 3 & N & G &   \\
\end{tabular}
\end{center}
\end{table*}

\begin{table*}
\begin{center}
Table 4---Continued \\
\begin{tabular}{l c r r c c c c} \colrule \colrule  
& & \multicolumn{1}{c}{$\mu$} & & & Distinct\tablenotemark{b} & Quality\tablenotemark{c} & 2~cm\tablenotemark{d} \\
\multicolumn{1}{c}{Source} & Comp. &
\multicolumn{1}{c}{(mas yr$^{-1}$)} & \multicolumn{1}{c}{$\beta_{app}$} &
$\xi$\tablenotemark{a}  & Component & Code & Survey \\ \colrule  
0920-397 & 1 & $0.867\pm0.436$  & $29.9\pm15.1$  & 2 & N & P & N \\
0923+392 & 1 & $0.122\pm0.119$  & $4.8\pm4.7$    & 2 & N & F & Y \\ 
         & 2 & $0.053\pm0.070$  & $2.1\pm2.8$    & 2 & Y & E &   \\ 
         & 3 & $0.044\pm0.107$  & $1.7\pm4.2$    & 2 & N & F &   \\ 
0953+254 & 1 & $-0.169\pm0.304$ & $-6.8\pm12.3$  & 2 & N & P & Y \\
         & 2 & $0.163\pm0.189$  & $6.6\pm7.6$    & 3 & N & P &   \\
0955+476 & 1 & $0.268\pm0.090$  & $21.5\pm7.2$   & 2 & N & F & N \\ 
         & 2 & $0.105\pm0.143$  & $8.4\pm11.4$   & 2 & N & F &   \\ 
1004+141 & 1 & $-0.060\pm0.034$ & $-5.8\pm3.3$   & 4 & Y & E & N \\
 	 & 2 & $-0.542\pm0.260$ & $-53.0\pm25.4$ & 2 & N & P &   \\
	 & 3 & $-0.081\pm0.118$ & $-8.0\pm11.5$  & 2 & N & F &   \\
 	 & 4 & $0.004\pm0.054$  & $0.4\pm5.3$    & 3 & N & G &   \\
1022+194 & 3 & $0.176\pm0.059$  & $8.0\pm2.7$    & 3 & Y & G & N \\
         & 4 & $0.150\pm0.061$  & $6.8\pm2.8$    & 3 & N & F &   \\
1034-293 & 1 & $1.457\pm0.292$  & $28.5\pm5.7$   & 2 & N & P & N \\
         & 2 & $1.215\pm0.187$  & $23.7\pm3.6$   & 4 & N & G &   \\
1044+719 & 1 & $0.160\pm0.091$  & $9.4\pm5.3$    & 2 & N & F & N \\
1101+384 & 1 & $-0.272\pm0.235$ & $-0.5\pm0.5$   & 2 & N & P & Y \\
         & 2 & $0.000\pm0.041$  & $0.0\pm0.1$    & 4 & N & G &   \\
         & 3 & $-0.067\pm0.068$ & $-0.1\pm0.1$   & 3 & N & G &   \\
1124-186 & 1 & $0.033\pm0.539$  & $1.8\pm29.5$   & 2 & N & P & N \\
         & 2 & $0.072\pm0.098$  & $4.0\pm5.4$    & 3 & N & F &   \\
1128+385 & 1 & $0.001\pm0.038$  & $0.1\pm2.9$    & 3 & N & G & Y \\ 
         & 2 & $0.014\pm0.016$  & $1.1\pm1.2$    & 5 & N & G &   \\ 
1144-379 & 1 & $-0.195\pm0.140$ & $-10.6\pm7.7$  & 3 & N & F & N \\
1145-071 & 1 & $0.059\pm0.012$  & $3.8\pm0.8$    & 5 & Y & E & N \\
1156+295 & 1 & $-0.386\pm0.274$ & $-15.9\pm11.3$ & 1 & N & P & Y \\
         & 2 & $0.452\pm0.185$  & $18.6\pm7.6$   & 1 & Y & P &   \\
         & 3 & $0.155\pm0.094$  & $6.4\pm3.9$    & 2 & N & F &   \\
1219+044 & 1 & $0.064\pm0.038$  & $3.3\pm2.0$    & 4 & N & G & N \\
1228+126 & 1 & $0.241\pm0.209$  & $0.07\pm0.06$  & 2 & N & P & Y \\
	 & 2 & $0.251\pm0.095$  & $0.07\pm0.03$  & 2 & N & F &   \\
         & 3 & $0.051\pm0.066$  & $0.01\pm0.02$  & 3 & N & G &   \\
1253-055 & 1 & $0.181\pm0.046$  & $5.7\pm1.5$    & 4 & Y & G & Y \\
         & 2 & $0.188\pm0.120$  & $6.0\pm3.8$    & 3 & N & P &   \\
         & 3 & $-0.027\pm0.120$ & $-0.9\pm3.8$   & 3 & N & P &   \\    
1255-316 & 2 & $-0.848\pm0.699$ & $-69.1\pm56.9$ & 3 & Y & P & N \\
	 & 3 & $0.248\pm0.561$  & $20.2\pm45.7$  & 3 & Y & P &   \\
1308+326 & 1 & $0.343\pm0.014$  & $18.0\pm0.8$   & 5 & N & G & Y \\ 
1313-333 & 1 & $0.376\pm0.083$  & $22.7\pm5.0$   & 3 & N & F & N \\
         & 2 & $0.156\pm0.056$  & $9.5\pm3.4$    & 3 & N & G &   \\ 
1334-127 & 1 & $0.011\pm0.038$  & $0.4\pm1.2$    & 5 & Y & E & Y \\ 
         & 2 & $0.276\pm0.080$  & $8.8\pm2.6$    & 4 & N & G &   \\ 
1351-018 & 1 & $0.278\pm0.250$  & $31.4\pm28.1$  & 3 & N & P & N \\
1418+546 & 1 & $0.055\pm0.079$  & $0.5\pm0.8$    & 3 & Y & E & N \\
         & 2 & $-0.008\pm0.079$ & $-0.1\pm0.8$   & 3 & Y & E &   \\
         & 3 & $0.027\pm0.198$  & $0.3\pm1.9$    & 2 & N & P &   \\
1424-418 & 1 & $-0.001\pm0.887$ & $-0.1\pm62.6$  & 2 & N & P & N \\
1451-375 & 1 & $-0.533\pm1.972$ & $-10.5\pm38.7$ & 1 & N & P & N \\
         & 2 & $-0.011\pm0.377$ & $-0.2\pm7.4$   & 2 & N & P &   \\
1514-241 & 1 & $3.593\pm0.746$  & $11.6\pm2.4$   & 1 & N & P & N \\
         & 2 & $0.957\pm0.360$  & $3.1\pm1.2$    & 2 & N & P &   \\
         & 3 & $0.337\pm0.337$  & $1.1\pm1.1$    & 2 & N & P &   \\
         & 4 & $0.231\pm0.192$  & $0.7\pm0.6$    & 3 & N & P &   \\
1606+106 & 1 & $-0.136\pm0.070$ & $-8.3\pm4.3$   & 3 & N & G & Y \\ 
         & 2 & $0.019\pm0.063$  & $1.1\pm3.8$    & 3 & N & G &   \\ 
         & 3 & $0.089\pm0.049$  & $5.5\pm3.0$    & 3 & N & G &   \\ 
         & 4 & $-0.092\pm0.032$ & $-5.6\pm2.0$   & 4 & N & G &   \\ 
1611+343 & 1 & $0.029\pm0.022$  & $1.9\pm1.5$    & 4 & N & G & Y \\
         & 2 & $0.063\pm0.022$  & $4.2\pm1.5$    & 4 & Y & E &   \\
         & 3 & $0.027\pm0.073$  & $1.8\pm4.9$    & 3 & N & G &   \\
1622-253 & 1 & $0.229\pm0.138$  & $10.0\pm6.0$   & 2 & N & F & N \\ 
         & 2 & $-0.130\pm0.274$ & $-5.7\pm12.0$  & 2 & N & P &   \\ 
1638+398 & 1 & $-0.022\pm0.077$ & $-1.7\pm5.8$   & 2 & N & G & N \\ 
\end{tabular}
\end{center}
\end{table*}

\begin{table*}
\begin{center}
Table 4---Continued \\
\begin{tabular}{l c r r c c c c} \colrule \colrule  
& & \multicolumn{1}{c}{$\mu$} & & & Distinct\tablenotemark{b} & Quality\tablenotemark{c} & 2~cm\tablenotemark{d} \\
\multicolumn{1}{c}{Source} & Comp. &
\multicolumn{1}{c}{(mas yr$^{-1}$)} & \multicolumn{1}{c}{$\beta_{app}$} &
$\xi$\tablenotemark{a}  & Component & Code & Survey \\ \colrule  
1652+398 & 1 & $-0.090\pm0.111$ & $-0.2\pm0.2$   & 2 & Y & F & Y \\
	 & 2 & $0.162\pm0.113$  & $0.4\pm0.3$    & 2 & Y & F &   \\
	 & 3 & $0.090\pm0.059$  & $0.2\pm0.1$    & 3 & Y & G &   \\
	 & 4 & $0.009\pm0.030$  & $0.0\pm0.1$    & 4 & N & F &   \\
1726+455 & 1 & $0.202\pm0.045$  & $8.2\pm1.8$    & 3 & Y & E & N \\
         & 2 & $0.025\pm0.127$  & $1.0\pm5.2$    & 3 & N & F &   \\
1739+522 & 1 & $-0.254\pm0.119$ & $-16.8\pm7.9$  & 1 & N & F & N \\
1741-038 & 3 & $0.004\pm0.528$  & $0.2\pm28.9$   & 2 & N & P & N \\ 
         & 4 & $0.035\pm0.056$  & $1.9\pm3.1$    & 3 & N & G &   \\ 
1745+624 & 1 & $-0.116\pm0.098$ & $-13.3\pm11.2$ & 2 & N & F & N \\
1749+096 & 1 & $0.782\pm0.202$  & $15.7\pm4.1$   & 1 & N & P & Y \\ 
         & 2 & $0.661\pm0.107$  & $13.3\pm2.1$   & 2 & N & G &   \\ 
         & 3 & $0.217\pm0.087$  & $4.4\pm1.7$    & 3 & N & F &   \\ 
         & 4 & $-0.254\pm0.936$ & $-5.1\pm18.8$  & 2 & N & P &   \\ 
1803+784 & 1 & $0.189\pm0.068$  & $7.3\pm2.6$    & 2 & N & G & Y \\
         & 2 & $0.119\pm0.076$  & $4.6\pm3.0$    & 2 & N & G &   \\
         & 3 & $0.078\pm0.073$  & $3.0\pm2.8$    & 2 & N & G &   \\
         & 4 & $0.027\pm0.017$  & $1.0\pm0.7$    & 4 & Y & E &   \\
         & 5 & $0.193\pm0.034$  & $7.5\pm1.3$    & 3 & N & G &   \\
         & 6 & $0.137\pm0.027$  & $5.3\pm1.0$    & 4 & N & G &   \\
1908-201 & 1 & $0.600\pm0.217$  & $34.3\pm12.4$  & 2 & N & P & N \\
         & 2 & $0.225\pm0.121$  & $12.9\pm6.9$   & 3 & Y & G &   \\
1921-293 & 1 & $0.241\pm0.110$  & $5.3\pm2.4$    & 2 & Y & G & Y \\ 
         & 2 & $-0.101\pm0.216$ & $-2.2\pm4.7$   & 2 & N & P &   \\ 
         & 3 & $0.333\pm0.234$  & $7.3\pm5.1$    & 3 & N & P &   \\
1928+738 & 1 & $0.971\pm0.217$  & $18.4\pm4.1$   & 2 & Y & P & Y \\
         & 2 & $0.547\pm0.223$  & $10.4\pm4.2$   & 2 & Y & P &   \\
         & 3 & $0.619\pm0.216$  & $11.7\pm4.1$   & 2 & Y & P &   \\
         & 5 & $0.439\pm0.229$  & $8.3\pm4.3$    & 2 & N & P &   \\
1954-388 & 1 & $0.164\pm0.410$  & $6.0\pm15.0$   & 2 & N & P & N \\
	 & 2 & $-0.033\pm0.051$ & $-1.2\pm1.8$   & 5 & N & G &   \\
1958-179 & 1 & $0.206\pm0.372$  & $7.7\pm13.9$   & 2 & N & P & N \\
2136+141 & 1 & $0.108\pm0.162$  & $10.0\pm15.0$  & 2 & Y & P & Y \\
         & 2 & $0.593\pm0.246$  & $54.9\pm22.8$  & 1 & N & P &   \\
         & 3 & $0.327\pm0.041$  & $30.3\pm3.8$   & 4 & N & G &   \\
         & 4 & $0.167\pm0.044$  & $15.5\pm4.1$   & 4 & N & G &   \\  
2145+067 & 1 & $-0.178\pm0.156$ & $-9.3\pm8.2$   & 1 & N & P & Y \\ 
         & 2 & $-0.064\pm0.051$ & $-3.4\pm2.7$   & 3 & N & G &   \\ 
         & 3 & $0.105\pm0.036$  & $5.5\pm1.9$    & 3 & N & G &   \\ 
         & 4 & $0.046\pm0.030$  & $2.4\pm1.6$    & 5 & N & G &   \\ 
2200+420 & 1 & $1.253\pm0.136$  & $5.7\pm0.6$    & 2 & N & G & Y \\
         & 2 & $1.288\pm0.128$  & $5.8\pm0.6$    & 2 & Y & E &   \\
         & 3 & $0.556\pm0.126$  & $2.5\pm0.6$    & 2 & N & F &   \\
         & 4 & $1.183\pm0.146$  & $5.4\pm0.7$    & 2 & Y & E &   \\
         & 5 & $1.294\pm0.469$  & $5.9\pm2.1$    & 2 & N & P &   \\
2230+114 & 1 & $0.215\pm0.639$  & $11.7\pm34.6$  & 1 & N & P & Y \\
         & 2 & $-0.135\pm0.264$ & $-7.3\pm14.3$  & 2 & N & P &   \\
         & 3 & $-0.151\pm0.525$ & $-8.2\pm28.5$  & 1 & Y & P &   \\
         & 4 & $0.448\pm0.533$  & $24.3\pm28.9$  & 1 & N & P &   \\
         & 5 & $0.391\pm0.139$  & $21.1\pm7.5$   & 4 & N & F &   \\
         & 6 & $0.265\pm0.130$  & $14.3\pm7.0$   & 4 & Y & G &   \\ 
2234+282 & 1 & $0.072\pm0.031$  & $3.2\pm1.4$    & 3 & N & G & Y \\ 
2243-123 & 1 & $0.006\pm0.060$  & $0.2\pm2.2$    & 4 & Y & E & Y \\
         & 2 & $0.118\pm0.033$  & $4.3\pm1.2$    & 5 & Y & E &   \\
         & 3 & $0.035\pm0.035$  & $1.3\pm1.3$    & 5 & N & G &   \\
2255-282 & 1 & $0.004\pm0.032$  & $0.2\pm1.6$    & 4 & Y & E & N \\
         & 2 & $-0.048\pm0.108$ & $-2.4\pm5.4$   & 3 & N & P &   \\ \colrule
\end{tabular}
\end{center}
$a$: Error bars on component positions were computed as a fraction $1/2^{\xi}$ of the beam size,
where $\xi$ was a function of the average flux of the component, see
$\S$~\ref{appspeed}.\\
$b$: Whether or not the component is a distinct feature. Y=Yes, N=No.\\
$c$: Overall quality code, see text for definitions.\\
$d$: Whether or not the source has an apparent speed measured
from the 2~cm survey (K04). Y=Yes, N=No.
\end{table*}

\begin{figure*}
\begin{center}
\includegraphics[scale=0.975,angle=180.0]{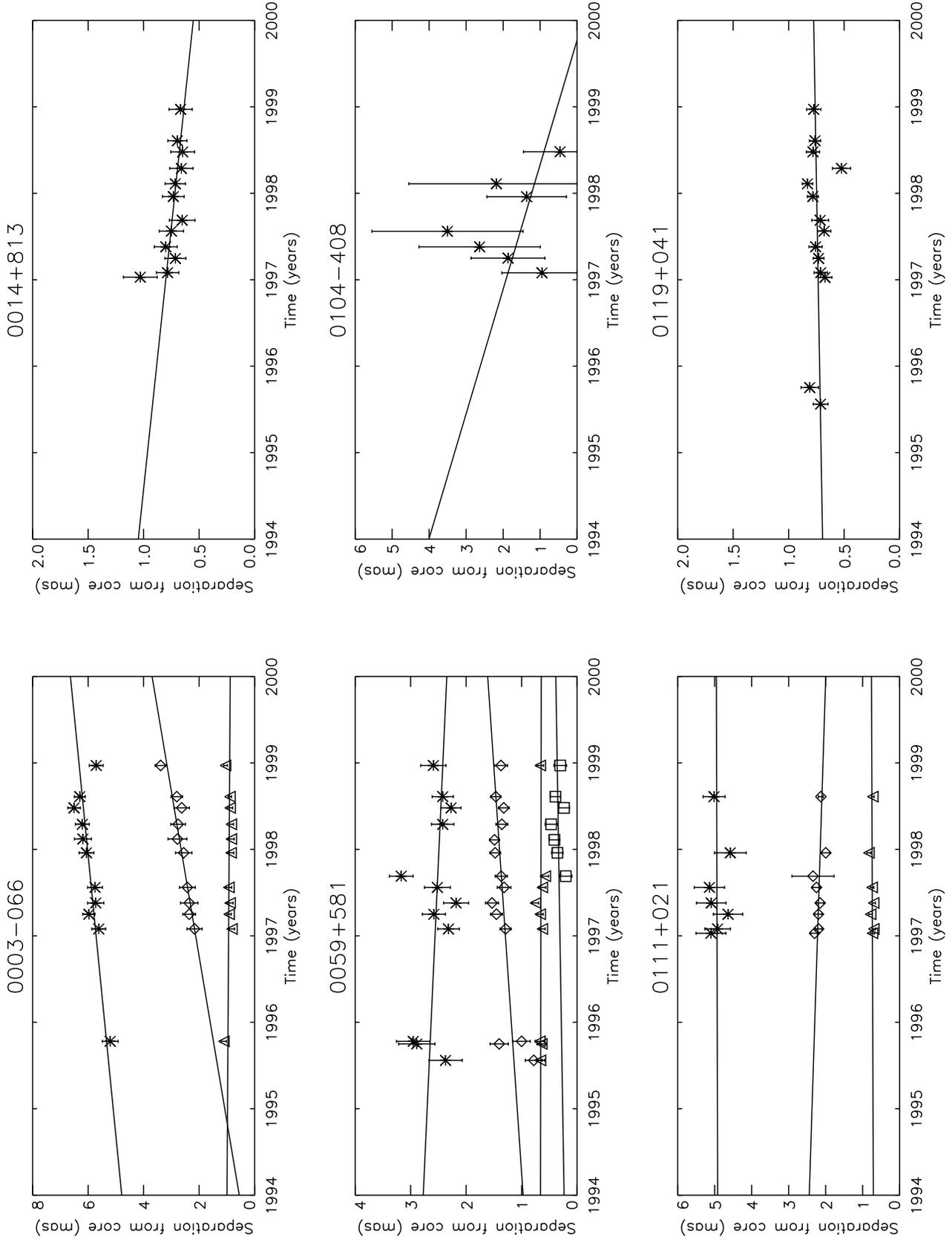}
\end{center}
\vspace{-0.40in}
\caption{Distances from the core of Gaussian component centers as
a function of time.
The lines are the least-squares fits to outward motion with constant speed.
For each source, asterisks are used to represent component 1, diamonds component 2, triangles component 3,
squares component 4, x's component 5, and circles component 6. Some error bars are smaller than the plotting symbols.}
\end{figure*}

\begin{figure*}
\begin{center}
\includegraphics[scale=0.975,angle=180.0]{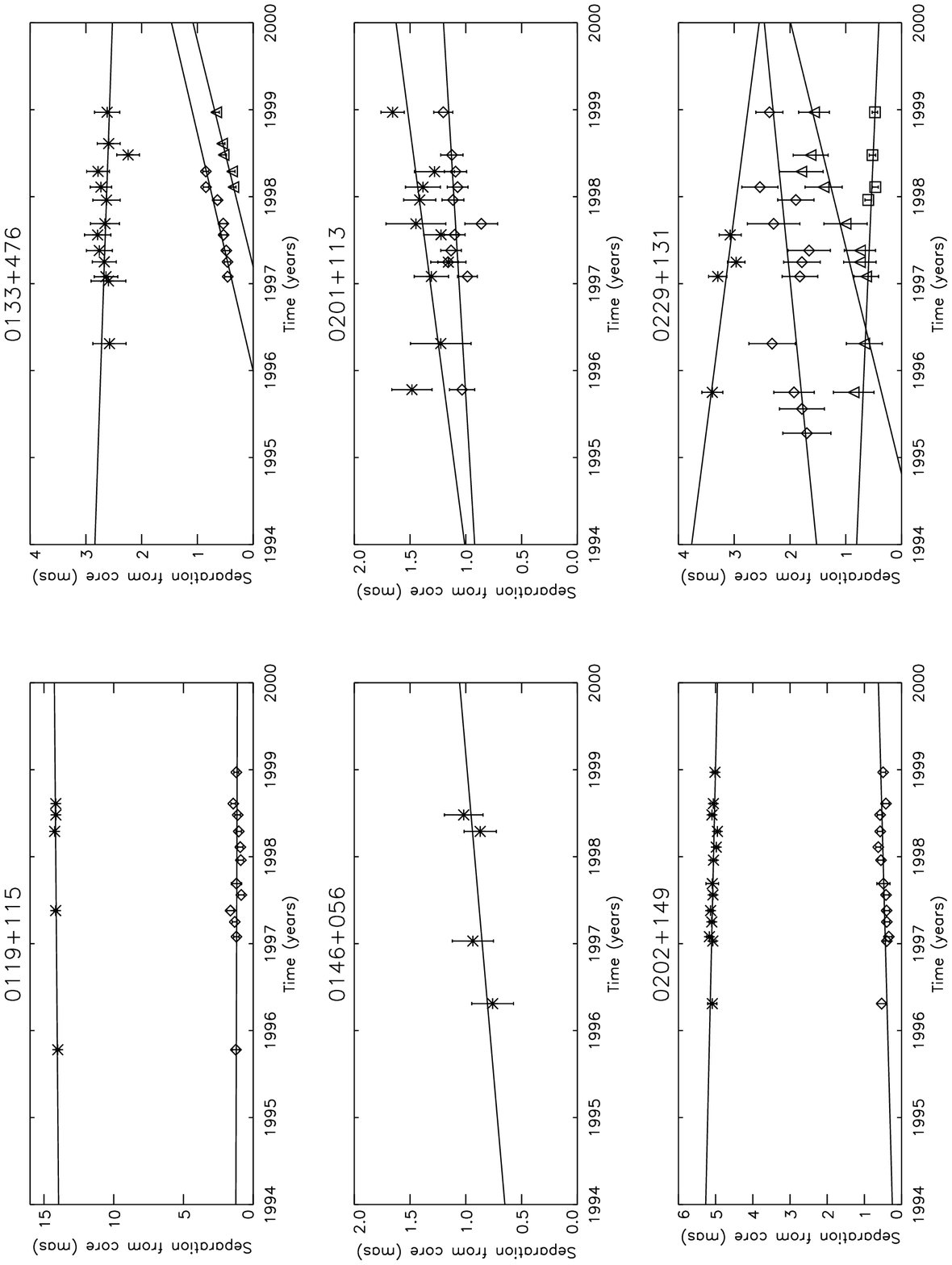}
FIG. 2.---{\em Continued}
\end{center}
\end{figure*}

\begin{figure*}
\begin{center}
\includegraphics[scale=0.975,angle=180.0]{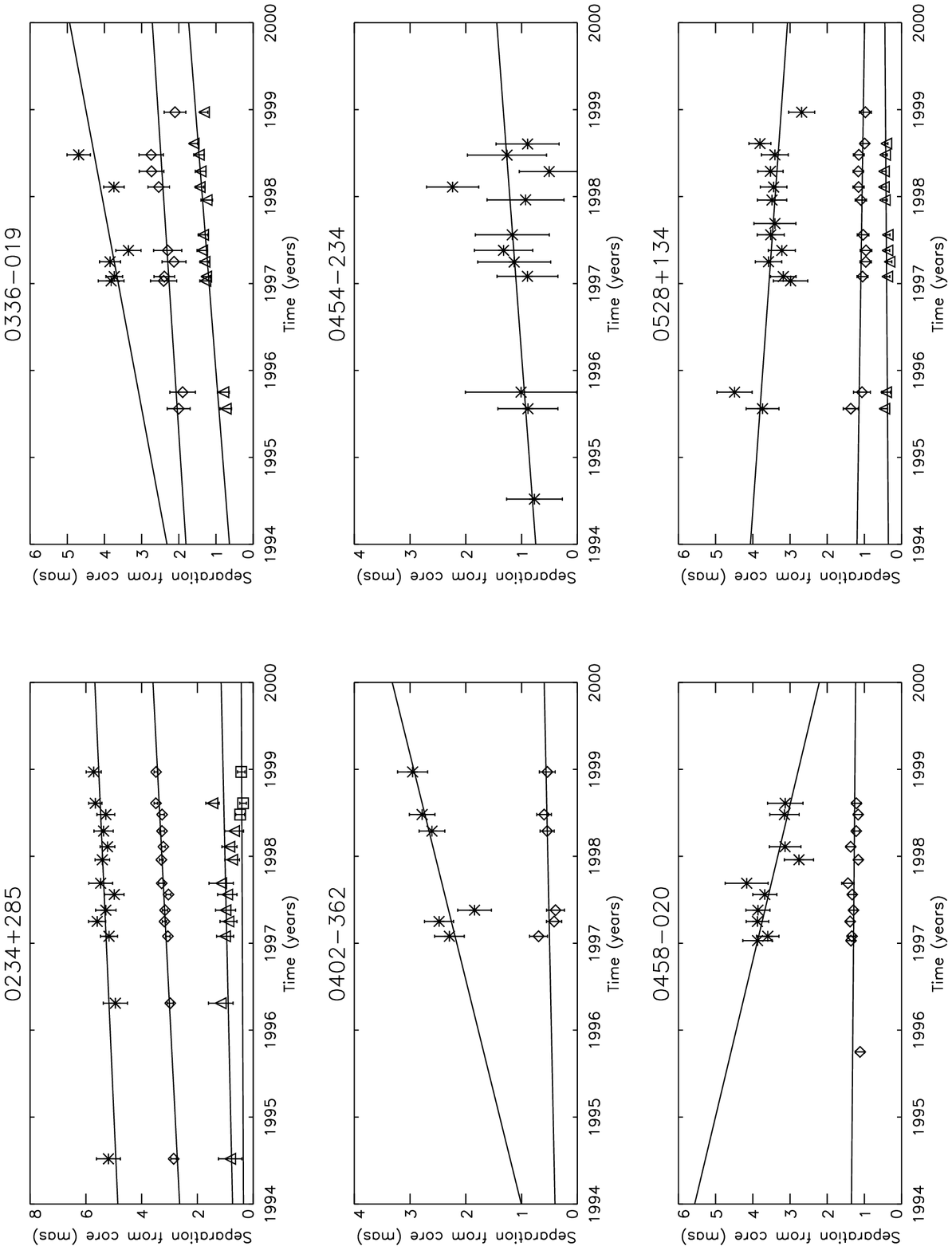}
FIG. 2.---{\em Continued}
\end{center}
\end{figure*}

\begin{figure*}
\begin{center}
\includegraphics[scale=0.975,angle=180.0]{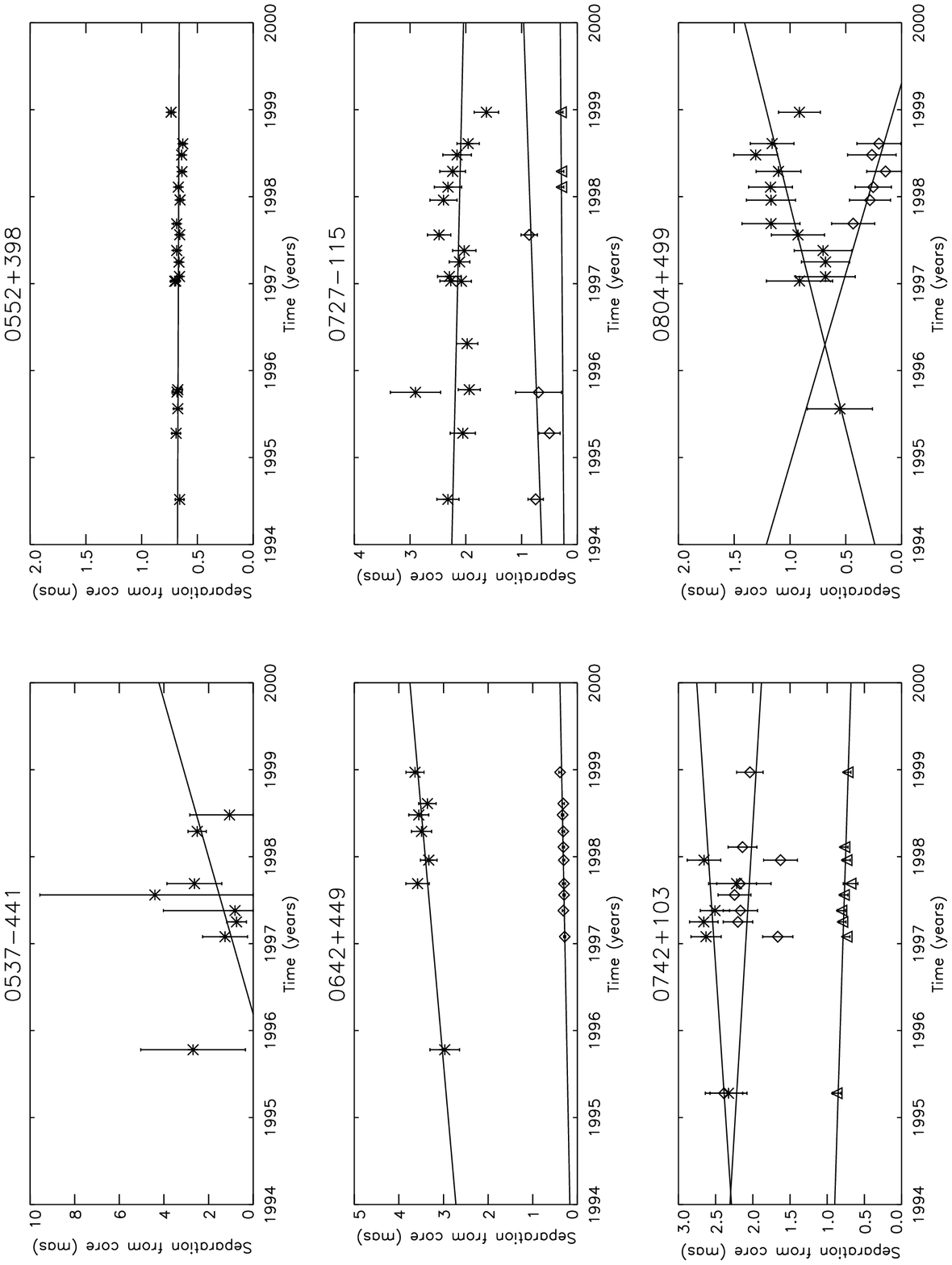}
FIG. 2.---{\em Continued}
\end{center}
\end{figure*}

\begin{figure*}
\begin{center}
\includegraphics[scale=0.975,angle=180.0]{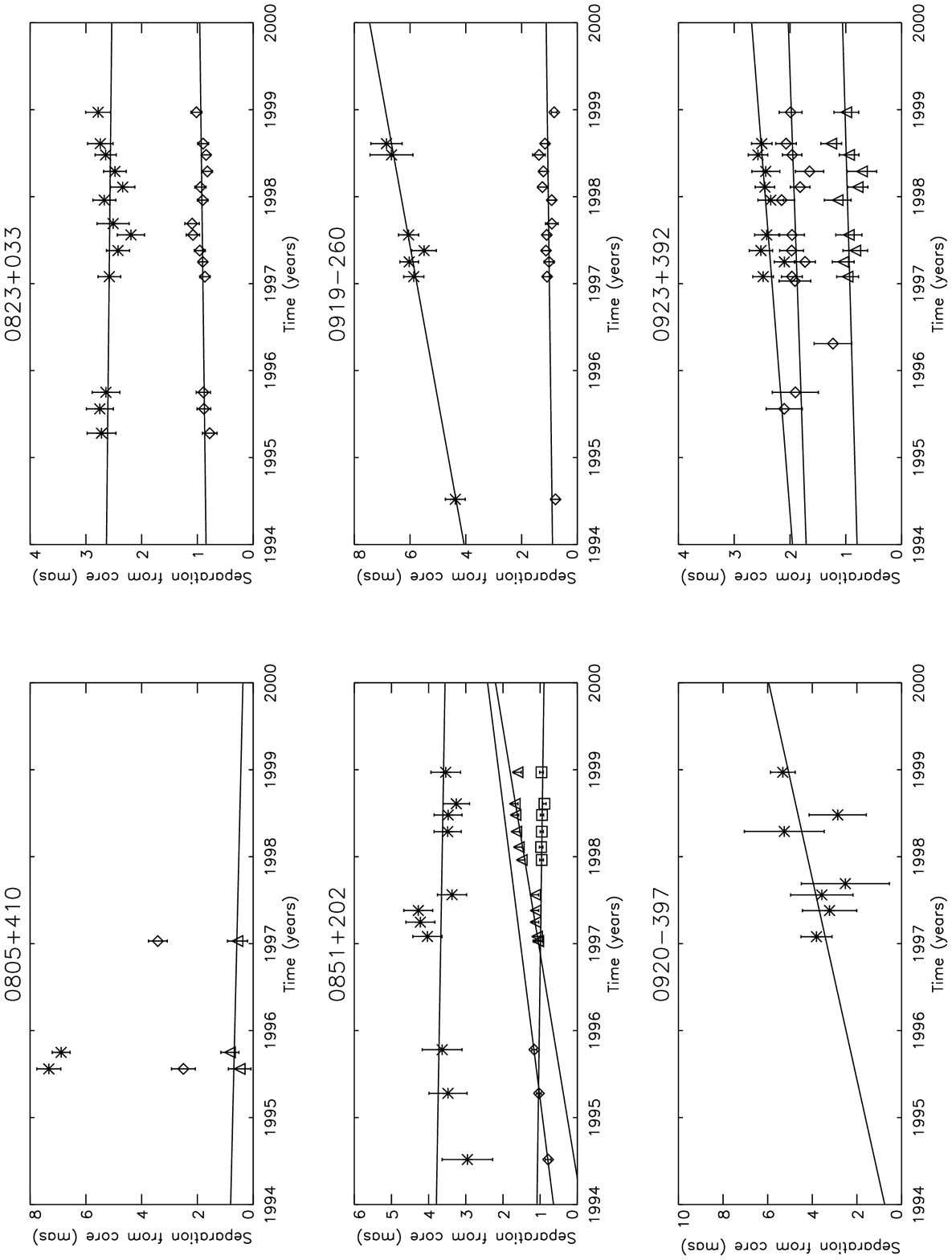}
FIG. 2.---{\em Continued}
\end{center}
\end{figure*}

\begin{figure*}
\begin{center}
\includegraphics[scale=0.975,angle=180.0]{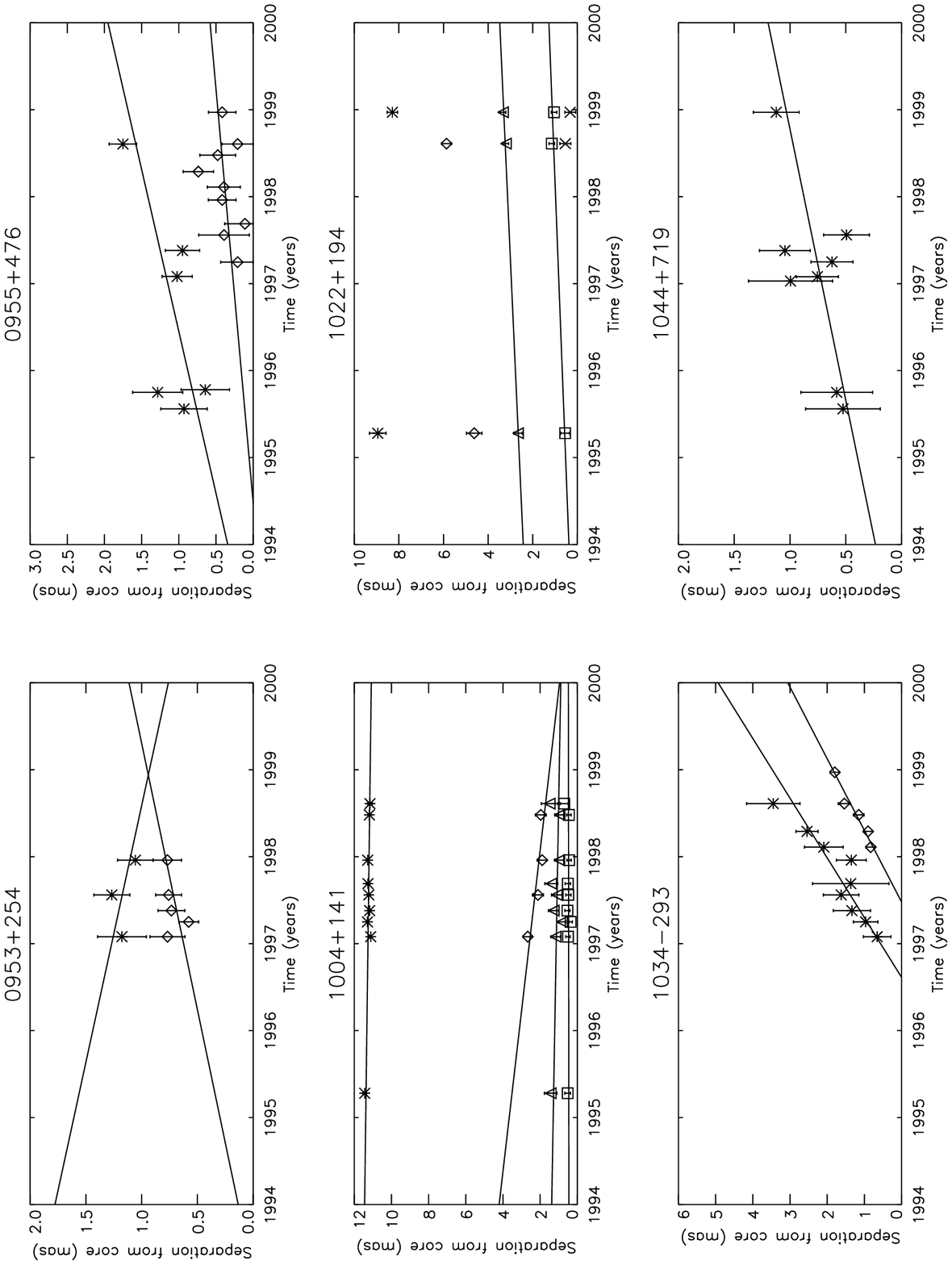}
FIG. 2.---{\em Continued}
\end{center}
\end{figure*}

\begin{figure*}
\begin{center}
\includegraphics[scale=0.975,angle=180.0]{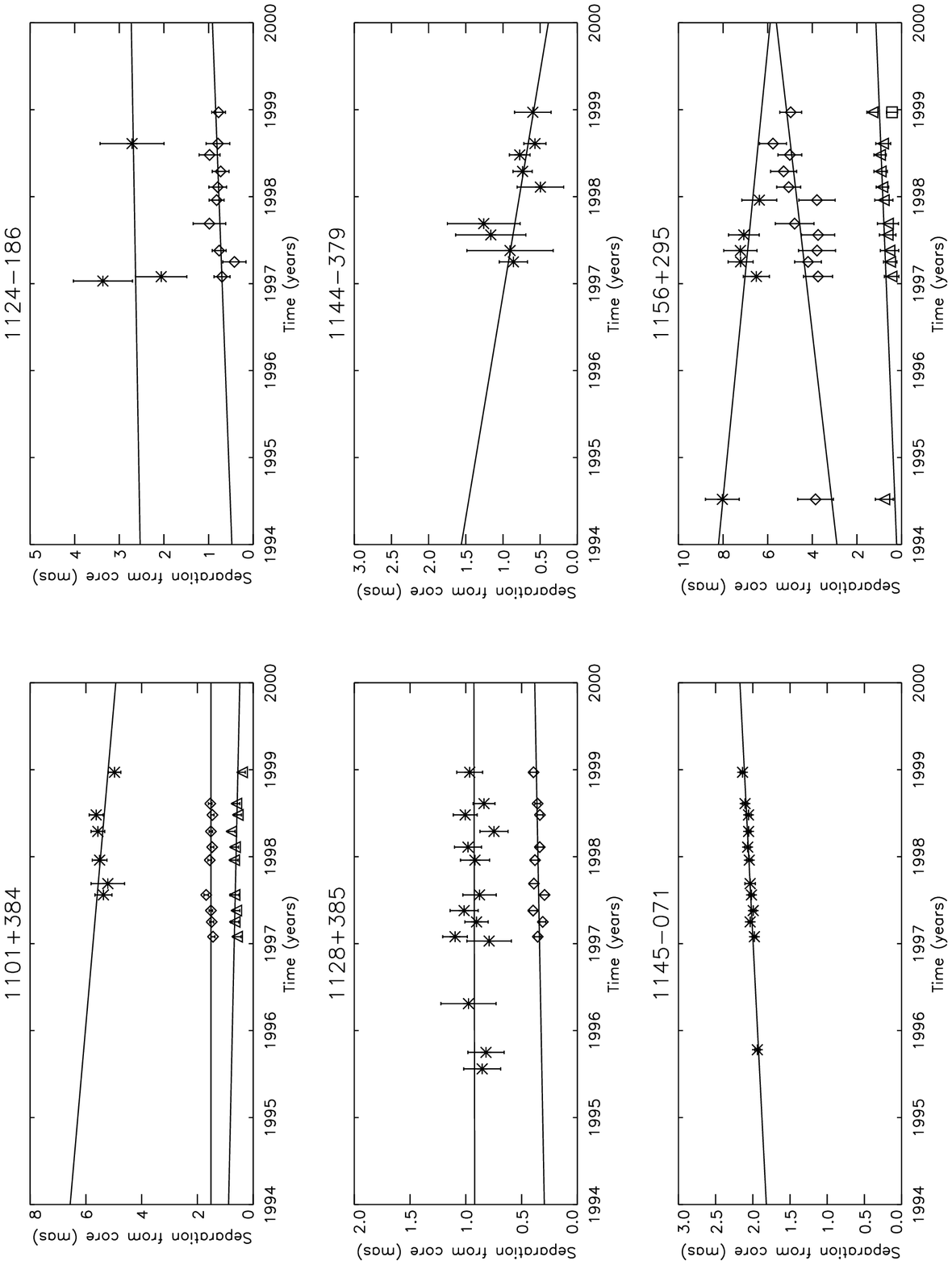}
FIG. 2.---{\em Continued}
\end{center}
\end{figure*}

\begin{figure*}
\begin{center}
\includegraphics[scale=0.975,angle=180.0]{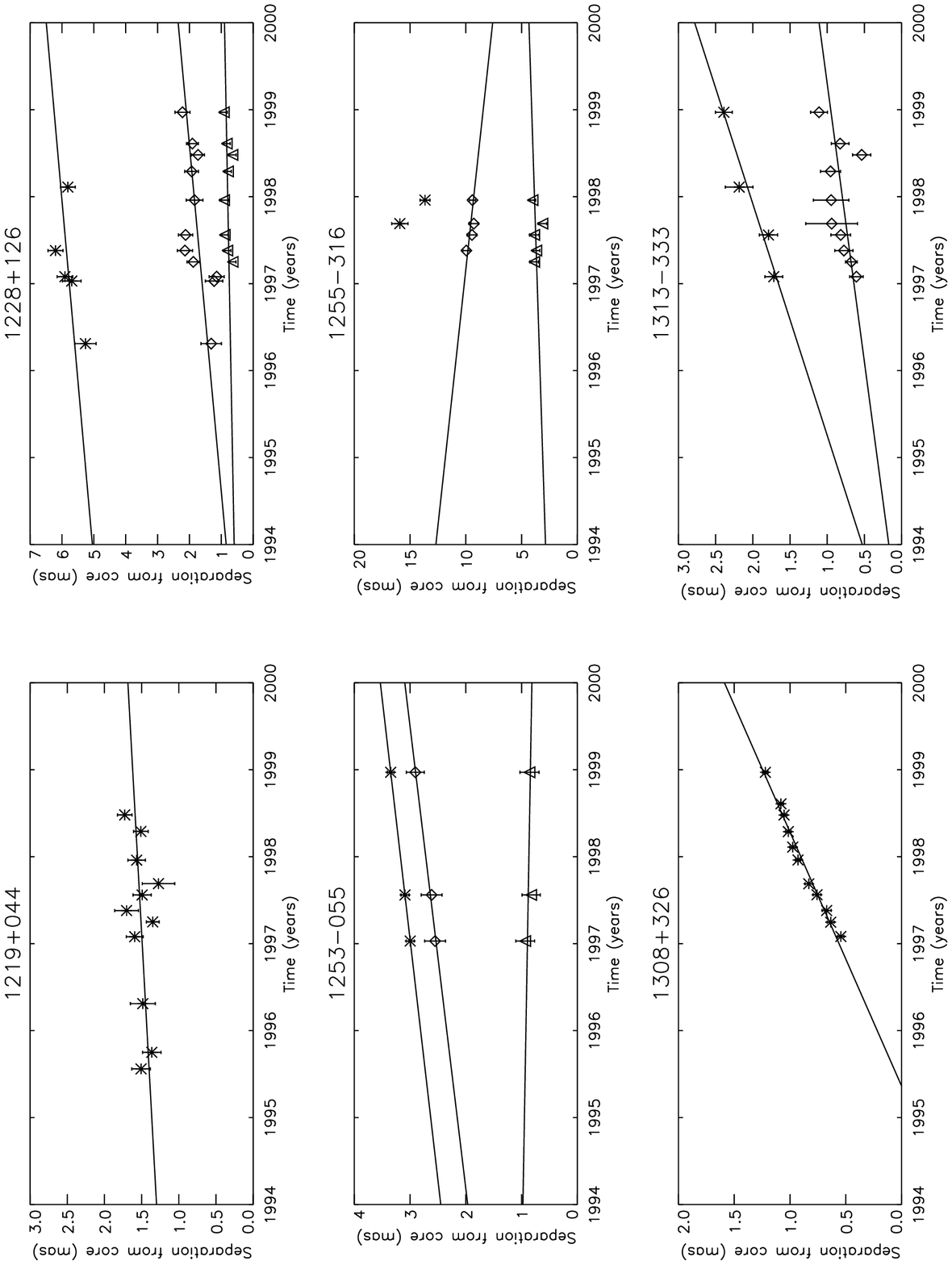}
FIG. 2.---{\em Continued}
\end{center}
\end{figure*}

\begin{figure*}
\begin{center}
\includegraphics[scale=0.975,angle=180.0]{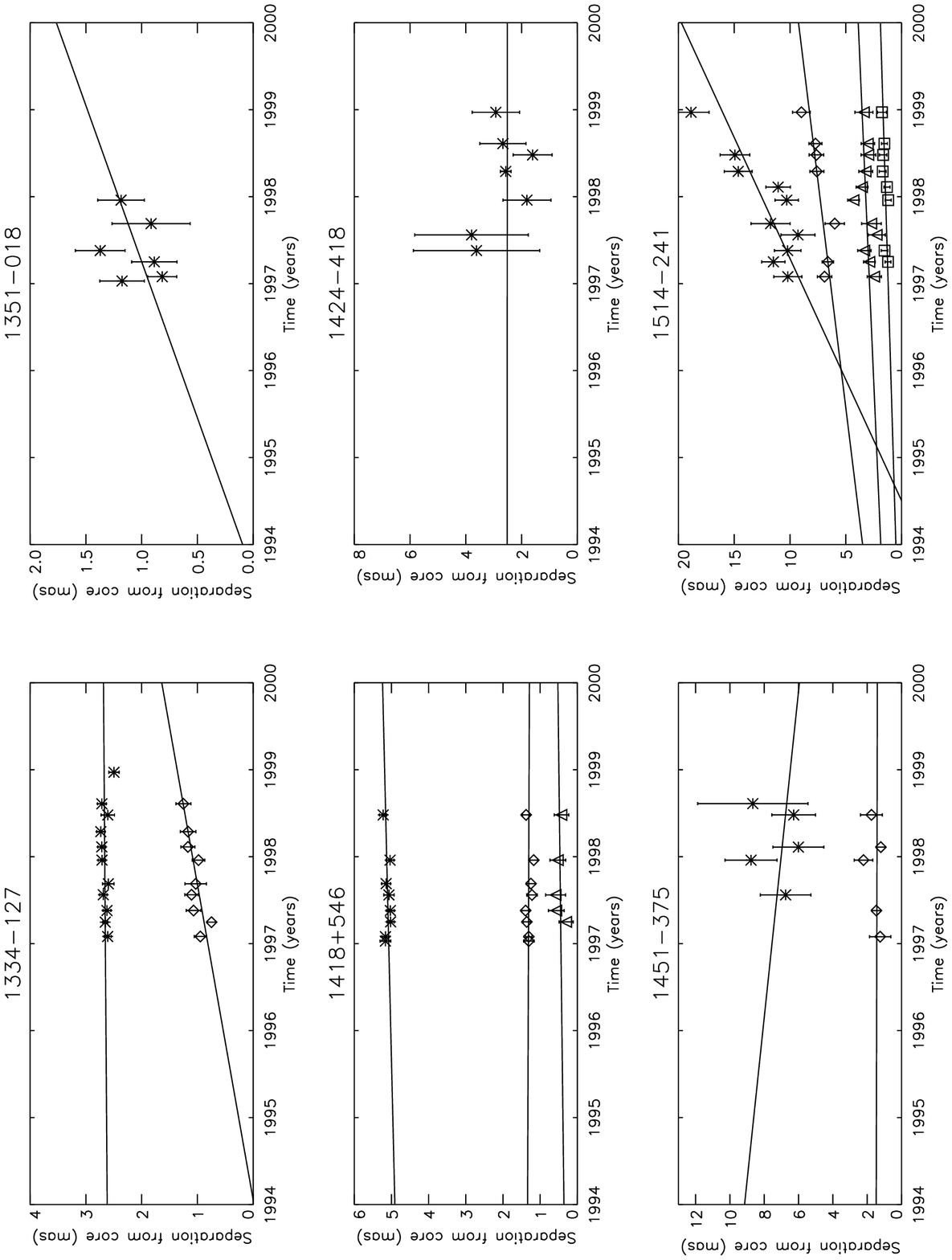}
FIG. 2.---{\em Continued}
\end{center}
\end{figure*}

\begin{figure*}
\begin{center}
\includegraphics[scale=0.975,angle=180.0]{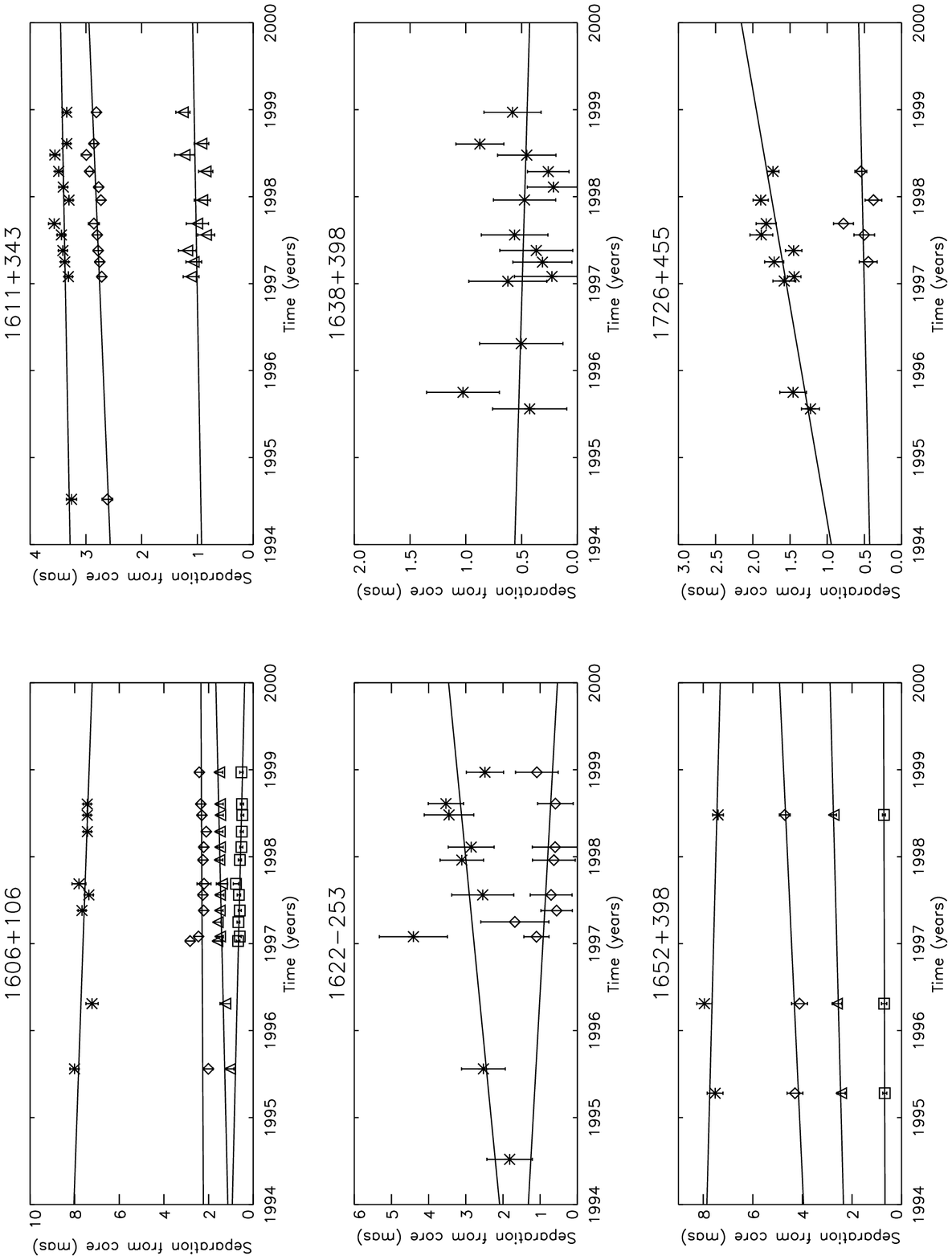}
FIG. 2.---{\em Continued}
\end{center}
\end{figure*}

\begin{figure*}
\begin{center}
\includegraphics[scale=0.975,angle=180.0]{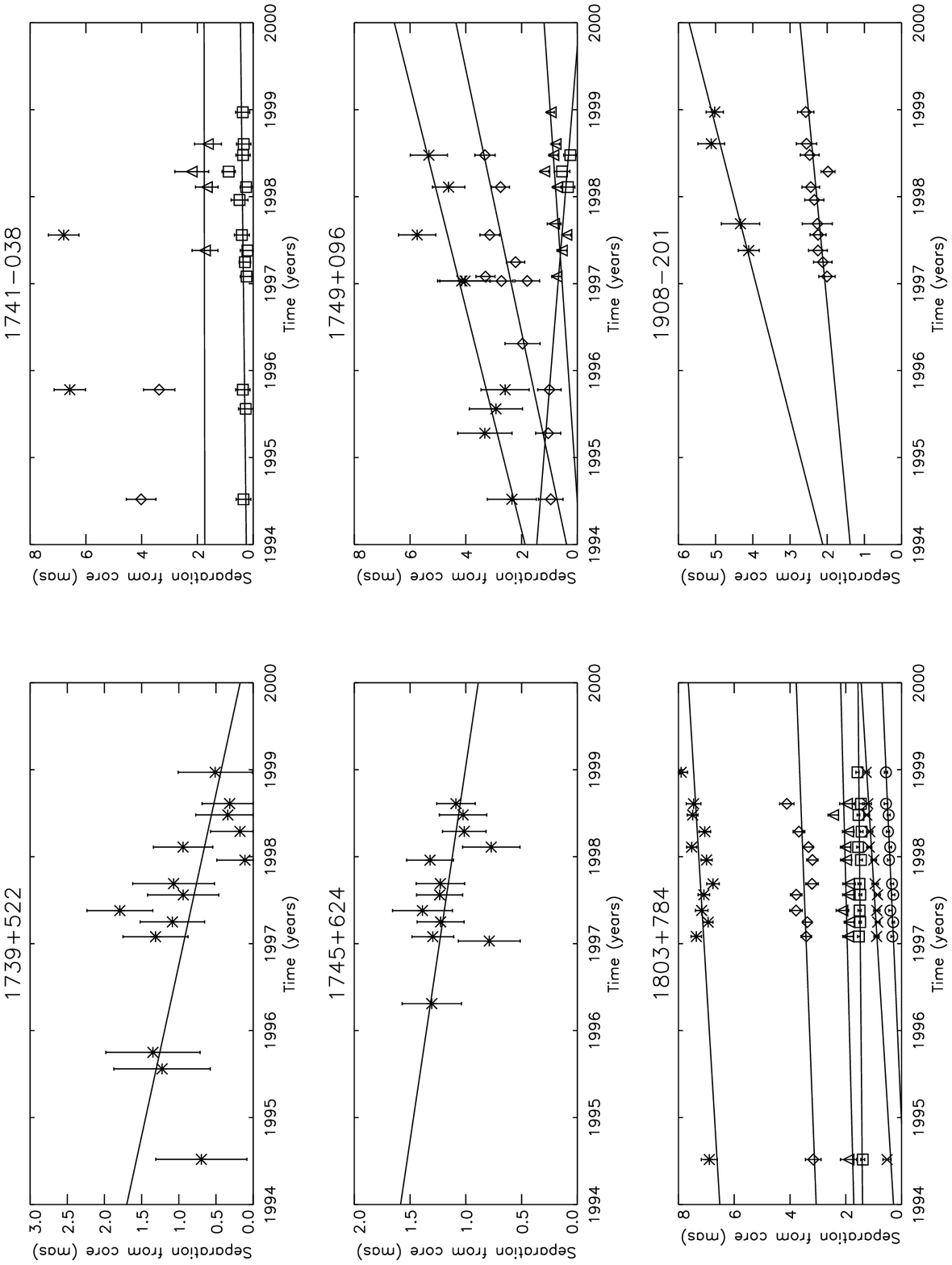}
FIG. 2.---{\em Continued}
\end{center}
\end{figure*}

\begin{figure*}
\begin{center}
\includegraphics[scale=0.975,angle=180.0]{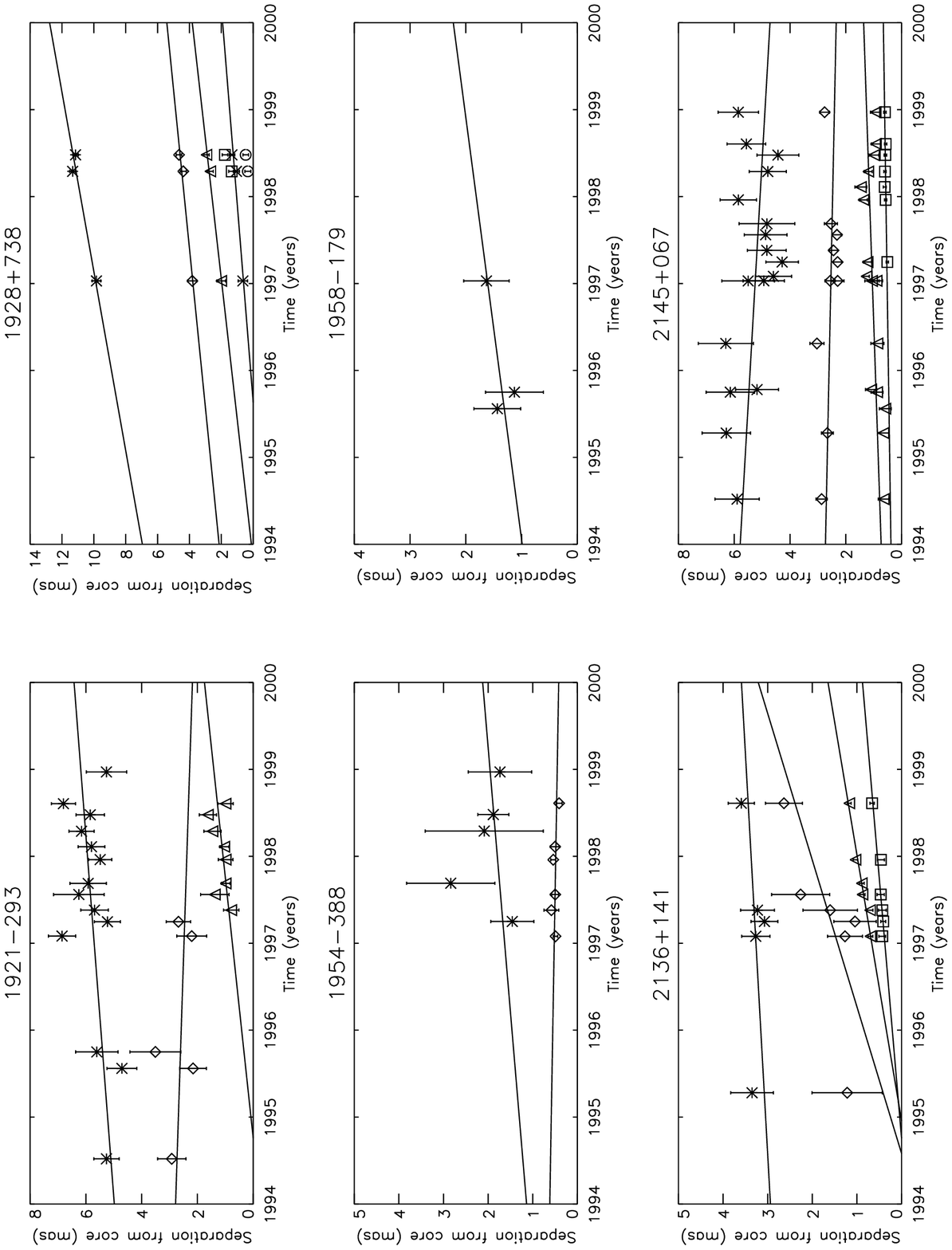}
FIG. 2.---{\em Continued}
\end{center}
\end{figure*}

\begin{figure*}
\begin{center}
\includegraphics[scale=0.975,angle=180.0]{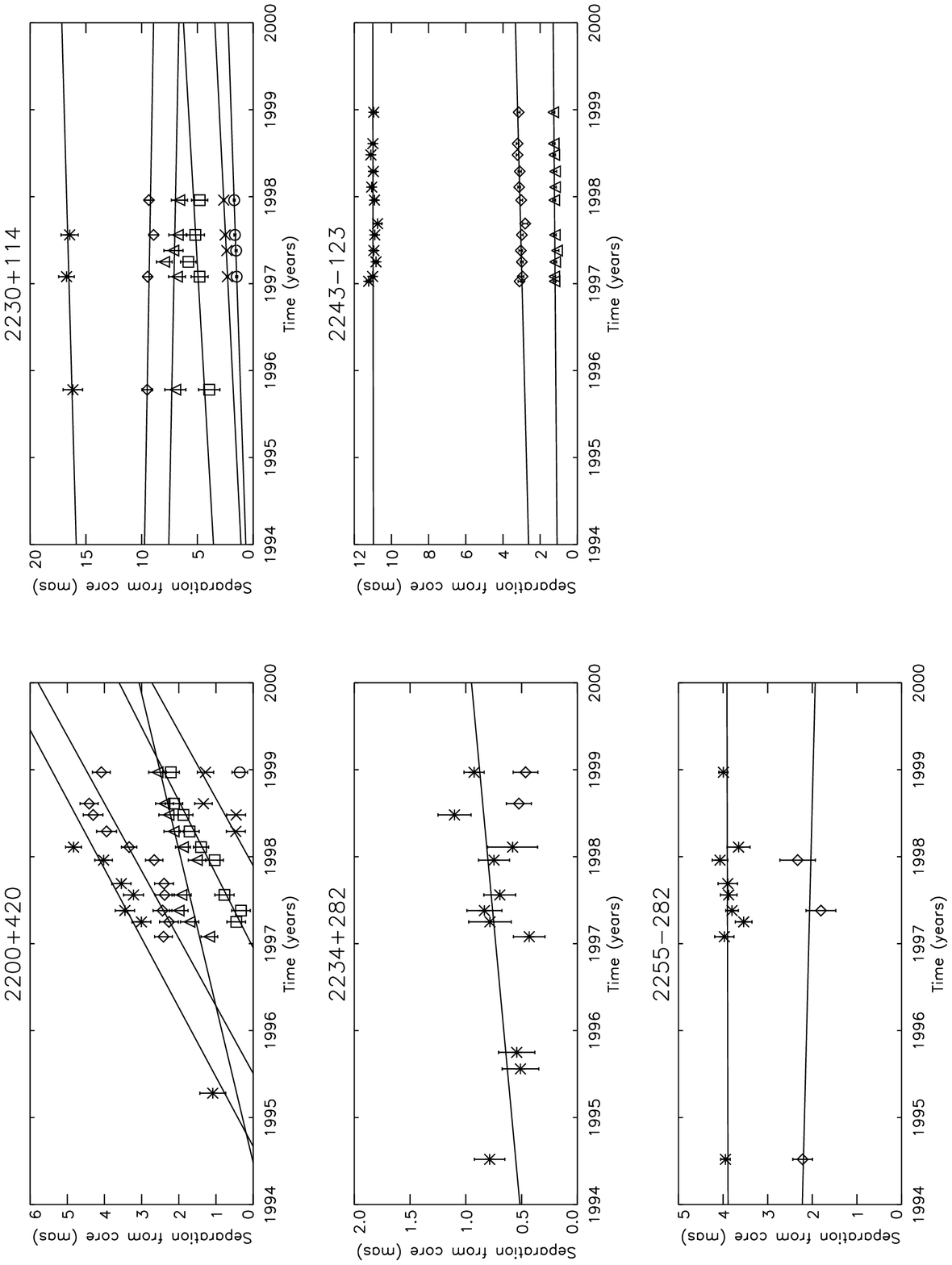}
FIG. 2.---{\em Continued}
\end{center}
\end{figure*}

Errors on the fitted radial positions of model components
were estimated by examining the scatter in the model-fit positions between sources present in the 
two pairs of adjacent epochs
1995OCT02-1995OCT12 and 1997JAN10-1997JAN11. A total of seven common sources with sixteen common components are present in
one or the other of these epoch pairs. The epochs comprising these two pairs are close enough together in
time that actual component motions are negligible, and any scatter in the
fitted positions represents the statistical errors in the model fits. 
Figure~3 shows a plot of the difference in the fitted radial positions of
the common components in these two epoch pairs (expressed as a fraction
of the beam size in the radial direction, because we expect the scatter to be proportional to the beam size)
versus the component flux (averaged between the two epochs). 
The positions of brighter components should be more accurately determined relative
to the beam size, and we expect a plot such as Figure~3 to show an upper envelope.
Figure~3 does appear to show such an envelope, and a fit of a power-law in flux to the eight points
near the upper envelope (asterisks in Figure~3) gives a fit close to $\Delta r\approx 2S^{-1/2}$, where
$\Delta r$ is measured in beams and $S$ is measured in mJy --- this curve is plotted
as a solid line on Figure~3. Error bars for individual components were estimated based on this curve
by setting the error bar size to be a fraction $1/2^{\xi}$ of the beam size, where $\xi$ was set from the
upper envelope fit given above as the closest integer to $1/2(\log S/\log 2)-1$, where $S$ is the
average flux of the component, and the maximum value of $\xi$ was 5.
The above procedure was used to set the default value for $\xi$, but since there are other factors that influence the
model-fitting accuracy, such as the
presence or absence of other confusing components, $\xi$ was adjusted from this default value if the error bars were
obviously way too large or too small (based on the significance of the fit), on a case-by-case basis.

We also computed all apparent speeds and associated errors a second time with no errors applied
to the individual position measurements, and the speed error calculated from the dispersion about the linear fit,
as was done for the 2~cm survey data by K04. The distributions of apparent speeds obtained by the two fitting methods
are statistically nearly identical according to a K-S test, and the two measured speeds for any given component
typically differ by much less than 1$\sigma$.
We retain the method described in the previous paragraph, because,
while the data in Figure~3 from which the flux dependence is derived is somewhat sparse, it at least takes
into account the changing resolutions from the early to the late epochs
(which went from VLBA-only to global VLBI, see Table~\ref{obstab}) by assigning beam-based errors,
and it is consistent with the values obtained from the dispersion about the linear fits.

An inspection of Table~4 shows a significant number of negative apparent speeds
(about 25\% of component speeds). A negative apparent
speed corresponds to inward motion toward the presumed core, and would be physically important if convincingly detected.
However, all of the negative apparent speeds in Table~4 are under 3$\sigma$ significance, and we
regard these as most likely due to stationary or slowly moving components that happen to have a formally negative
best-fit apparent speed. In fact, when we look at the distribution of significances of the negative apparent
speeds, they match closely a normal distribution centered at zero
(34\% have over 1$\sigma$ significance, 8\% have over 2$\sigma$ significance), supporting the interpretation in terms
of stationary or slowly moving components. 
Other VLBI surveys have reached similar conclusions regarding components with negative apparent speeds,
see, for example, the discussions by K04 and Vermeulen et al. (2003).

\begin{figure*}[!t]
\begin{center}
\includegraphics[scale=0.60]{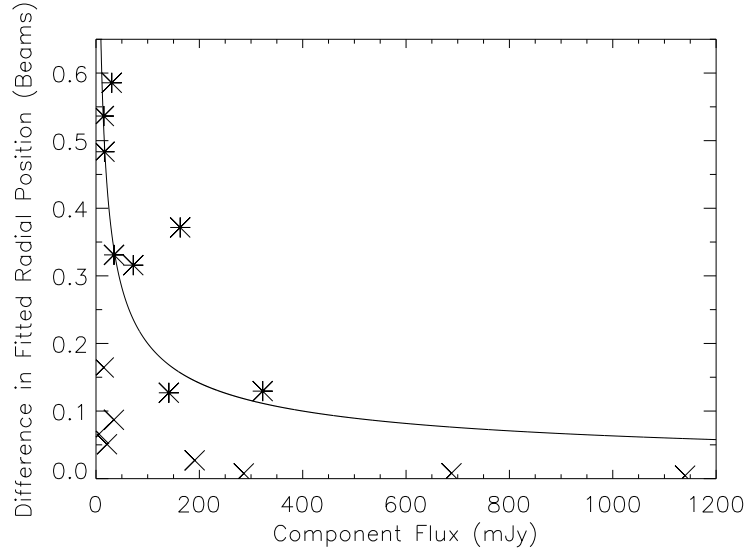}
\end{center}
\vspace{-0.20in}
\caption{Difference in the model-fit radial positions of common components in the adjacent epoch
pairs defined in $\S$~\ref{appspeed} versus the component flux. The beam size is computed as the
average of the radial beam size in the two epochs (projection of the beam onto a radial line
at the component's position angle), and the flux is the average flux from the two epochs. The solid
line shows an estimate of the upper envelope of this distribution, $\Delta r\approx 2S^{-1/2}$, where 
$\Delta r$ is measured in beams and $S$ is measured in mJy. The upper envelope was obtained by a fit to the eight
points indicated by asterisks, the points falling considerably below the upper envelope are indicated by X's.}
\end{figure*}

Note that there are some apparent speeds listed in Table~4 with extremely large associated errors
(e.g., the two components in 1255-316, with formal errors on the measured apparent speeds of about 50$c$). 
This has usually occurred because the RRFID is not a survey that was specifically designed to measure
jet kinematics, so occasionally a source may be observed for only a few epochs over a period of a few months.
Observing a component over a period of only a few months at 8 GHz yields large errors on the measured apparent speeds.
We include these components in Table~4 for completeness, and to show the current state of the RRFID observations
of these sources, but we caution that there is little or no information contained in those particular speed measurements, 
so that {\em they should not be used by
anyone for any reason}. 

To aid in the comparison to the 2~cm survey presented in $\S$~\ref{comp}, we assigned quality codes
to each component motion using the same criteria used by K04. These criteria are:
\begin{enumerate}
\item{The component is observed at four or more epochs.}
\item{The component is a well-defined feature in the images.}
\item{The uncertainty in the fitted proper motion is $\leq0.08$ mas yr$^{-1}$, or the proper motion has a significance
$\geq5\sigma$.}
\end{enumerate}
The quality codes are then assigned as follows:
\begin{enumerate}
\item{E (Excellent) for motions that satisfy all three of the above criteria.}
\item{G (Good) for motions that satisfy any two of the above criteria.}
\item{F (Fair) for motions that satisfy only one of the above criteria.}
\item{P (Poor) for motions that do not satisfy any of the above criteria, or for motions where
the uncertainty in the fitted proper motion is $>0.15$ mas yr$^{-1}$ (except for the $\geq5\sigma$
cases mentioned above).}
\end{enumerate}
These quality codes are listed in Table~4. Following K04, we restrict subsequent analysis
to those components having a Good or Excellent quality code.
This selection criterion excludes 90 of the 184 measured apparent speeds
in Table~4, leaving a total of 94 apparent speeds in 54 sources that are used in the subsequent analysis.
We note that because of the difference between the time baseline of the currently analyzed RRFID data (5 years) 
and that of the 2~cm survey (8 years), our typical uncertainty in fitted proper motions is about a factor of 8/5 larger
than the typical uncertainty from K04, causing a smaller yield of Excellent and Good components (about 
50\% of our components) compared to K04 (about 75\% of their components), due to the application of criterion (3) above.
These uncertainties in fitted proper motions will be reduced as more of the RRFID data is analyzed.

A histogram of the apparent speeds of the 94 `Good' and `Excellent' components is shown in Figure~4.
The mean apparent speed of these components is 3.6$c$.
The general shape of the distribution is similar to that found by other VLBI surveys, with a peak
at the lowest apparent speeds, and a tail extending out to higher speeds, up to
about 30$c$ in the case of Figure~4.  About half of the components are either in the lowest-speed 
or negative speed bins, and these are consistent with 
being stationary components.  Such stationary components are also observed to be common in the other large
multi-epoch VLBI surveys, including the radio-selected  
2~cm and CJF surveys (K04; Vermeulen 1995), and the VLBI survey of EGRET blazars by Jorstad et al. (2001).
In each of those surveys, one-third to one-half of the VLBI components observed were found to be slow or stationary
($<2c$ apparent speed when expressed in the cosmology given in $\S$\ref{intro}).  Numerical simulations of
relativistic jets also produce such stationary components;
for example, the simulations of relativistic jets by G\'{o}mez et al. (1995) suggested that such stationary components may
be standing oblique shocks created by quasi-periodic recollimation shocks.
Further simulations by Agudo et al. (2001) suggested that multiple slowly-moving conical shocks formed
from the interaction of superluminal components with the underlying jet could also produce
apparently stationary components on VLBI maps.

\begin{figure*}
\begin{center}
\includegraphics[scale=0.60]{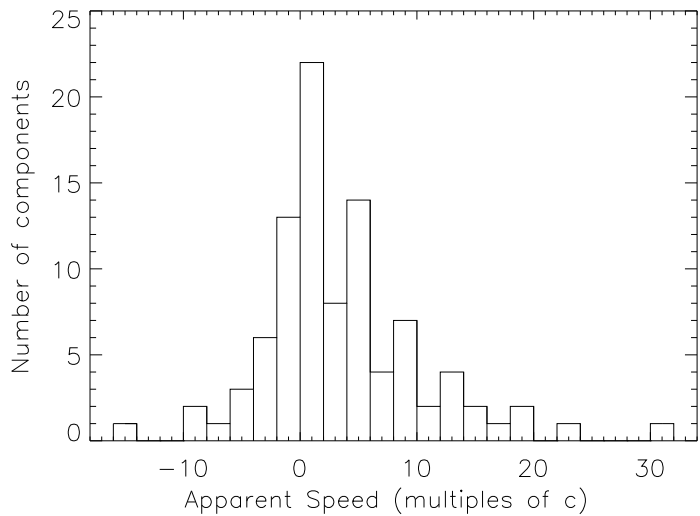}
\end{center}
\vspace{-0.20in}
\caption{Histogram of apparent component speeds for the 94 components
in Table~4 that have a `Good' or `Excellent' quality code.}
\end{figure*}


We have quantitatively compared the distribution shown in Figure~4 with the 
apparent speed distributions obtained by other authors using the Kolmogorov-Smirnov (K-S) test.
A direct comparison of the histogram in Figure~4 with the corresponding histogram
for the 2~cm survey from K04 shows a difference at the 95\% significance level; however, this difference
is solely the result of the greater scatter toward negative apparent speeds 
of the presumably stationary components caused
by the approximately 8/5 greater statistical uncertainty in the RRFID proper motions mentioned above.
If the negative apparent speeds from both surveys are grouped in their respective $0c$ bins,
then the K-S test finds no significant difference between the distribution of apparent speeds
measured here and that measured by the 2~cm survey.
Note that even though the general shapes of the overall apparent speed distributions agree quite well
between this paper and K04, specific results for some individual sources shared by the two surveys do 
differ, see $\S$\ref{comp} for a full discussion.
A K-S test shows a significant difference between the distribution in Figure~4 and the distribution of apparent
speeds in EGRET-selected gamma-ray blazars measured by Jorstad et al. (2001) with $>99.7\%$ confidence
(regardless of what is done with the negative apparent speeds).
This type of difference is expected if the gamma-ray emission is more highly beamed than the radio emission;
however, see the further discussion of this issue in $\S$\ref{gamma}.

Different components within the same jet can have different apparent speeds, as obviously demonstrated by the
co-existence of fast moving and stationary components in the same source.
Approximately one-third of the sources that have multiple components in Figure~4
have apparent speeds that are different from each other at greater than 99\% confidence.
However, the dispersion of apparent speeds for individual sources is typically less than that for the sample as a whole,
as was also found by K04 for the 2~cm survey.
For the 27 sources that have multiple components in Figure~4, all but five have a dispersion of their apparent speeds
that is less than the dispersion of all components in Figure~4 taken together ($6.9c$).
This demonstrates the existence of a `characteristic speed' associated with individual jets.
Analysis of correlations between the fastest observed apparent speed in a source and other source properties
using the 2~cm and MOJAVE survey samples (Lister 2006) has suggested that the fastest observed pattern speed
in a source is a good indicator of bulk apparent speed.
The `characteristic speed' mentioned above may then be the bulk flow speed of the jet, with individual components
moving at pattern speeds ranging from zero up to this bulk flow speed. A scenario such as this would be consistent
with jet simulations such as those of Agudo et al. (2001), where primary disturbances moving at the jet flow speed spawn
secondary `components' that move at slower pattern speeds (or are stationary).
In this interpretation, the maximum apparent speed in a source, measured over a 
time interval of many years, would be the most reliable indicator of bulk Lorentz factor.
In Figure~5, we show the distribution of maximum apparent speeds for the RRFID sample.
The figure shows a histogram of the fastest measured pattern speed in each of the 54 sources represented in Figure~4.
The mean fastest apparent speed in a source is $5.9c$.
Note that an excess of stationary components could still be produced in such a plot if some sources had not yet been observed
over a long enough time interval to see a component moving at the apparent bulk speed.

\begin{figure*}
\begin{center}
\includegraphics[scale=0.60]{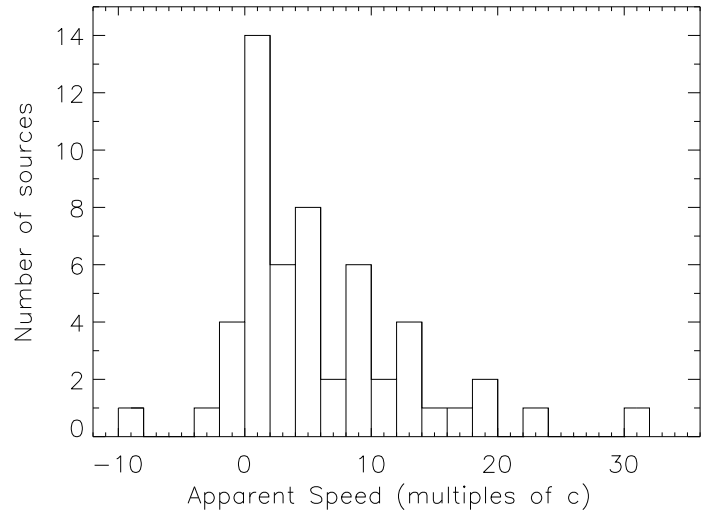}
\end{center}
\vspace{-0.20in}
\caption{Histogram of the fastest apparent component speed in each source,
for the 54 sources represented in Figure~4.} 
\end{figure*}

Figure~6 shows the distribution in Figure~5 separated by the optical type given in Table~\ref{sources} into quasars
and BL Lac objects. (Because only three sources in Figure~5 are classified as galaxies, we do not show a separate histogram
for the galaxies.) 
The mean fastest apparent speed for the quasars is $6.8\pm1.1c$, and that for the BL Lac objects is $3.2\pm1.5c$.
This difference in the means is significant at the 94\% confidence level.
For comparison, K04 found that their observed apparent speed distributions for quasars and BL Lac objects
differed at the 98\% confidence level, and Jorstad et al. (2001) found that the apparent speed distributions of
EGRET-selected quasars and EGRET-selected BL Lac objects differed at the 99.9\% confidence level.
Why this difference should be statistically more significant for the groups of gamma-ray selected quasars
and BL Lac objects than for the radio-selected groups is unclear.

\begin{figure*}
\begin{center}
\includegraphics[scale=0.50]{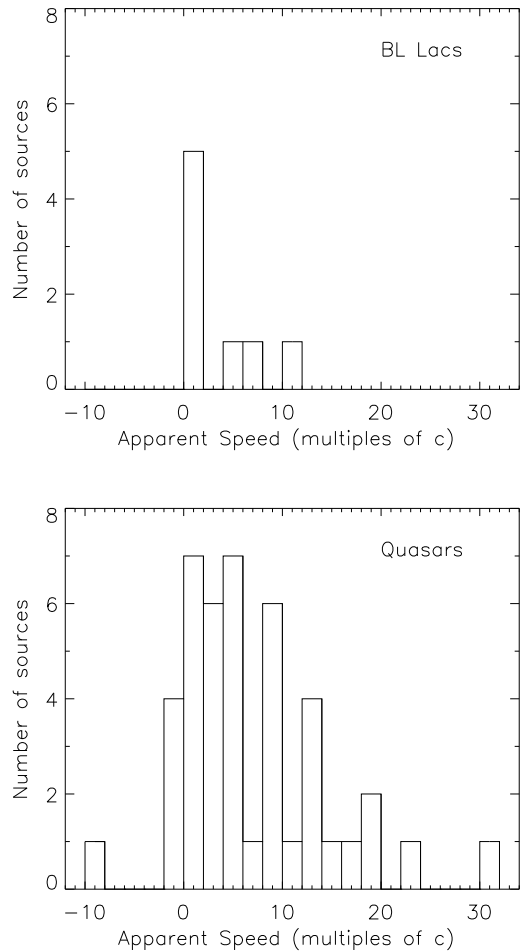}
\end{center}
\vspace{-0.20in}
\caption{Histograms of the fastest apparent component speed in each source, for the 54 sources represented
in Figure~5, separated by optical type. The upper panel shows the BL Lac objects (8 sources),
and the bottom panel shows the quasars (43 sources). Three sources classified as galaxies are not shown.}
\end{figure*}

\subsection{Apparent Nonradial Motions and Accelerations}
In this subsection, we check for two different types of apparent accelerations in the modeling data.
First, we perform second-order fits to $r$ versus $t$ (three free parameters), 
to check for apparent radial accelerations or
decelerations that could be caused by a changing Lorentz factor, angle to the line-of-sight, or direction of motion.
The thirteen closely spaced epochs during 1997 and 1998 should give some sensitivity to such second-order
terms in the apparent radial motion.  Secondly, we perform first order fits to $x$ versus $t$ and $y$ versus $t$
(four free parameters), which allows us to fit for a direction of motion of the component
that may be different from its mean position angle, or nonradial.
With the current time baseline and resolution of the RRFID data, higher-order
fits with more free parameters do not give meaningful results, although they will as the time
baseline of the RRFID continues to be extended in future papers.

We fit all components that were observed at four or more epochs (161 components) with a second-order
polynomial for $r(t)$. To check for significant accelerations, we considered all components where the second-order
term had greater than 2$\sigma$ significance. 
There were 17 such components, where $\approx8$ such $2\sigma$ detections are expected by chance alone,
suggesting some detections of real accelerations.
From those 17 components, we exclude those with `Fair' or `Poor' quality codes, and those
where the extremum of the second-order function occurs within
the time range of the component observations. Such fits would represent a component that reverses its direction
of motion during the observed time interval --- we consider this physically unlikely, and expect that these results
are caused by a statistically abnormally high or low point or points in the midst of the component position data.
There remain three components that we consider to be the most likely detections of radial accelerations: component C2 in 0234+285
($a=0.059\pm0.028$ mas yr$^{-2}$), component C1 in 1308+326 ($a=-0.130\pm0.051$ mas yr$^{-2}$), and component
C1 in 2200+420 ($a=0.717\pm0.295$ mas yr$^{-2}$). Two of these accelerations are positive and one is negative.
A positive apparent radial acceleration could be due to an increase in the bulk Lorentz factor of the
component (as is suggested to occur on parsec scales in some models, such as the accelerating MHD jet model of
Vlahakis \& K\"{o}nigl 2004), or to a bend toward the critical viewing angle for maximum apparent speed.
Negative apparent radial accelerations could be caused by a decrease in the bulk Lorentz factor, a bend away
from the critical angle, or a curved trajectory that acquires a significant nonradial component.
In fact, two of these three sources do have detectable nonradial motions, as discussed below.
These various scenarios will also produce variations in the apparent brightness of the component as the beaming factor changes,
but such variations are complicated by not knowing the true viewing angle or intrinsic variability of the component.

The second-order radial fits discussed above are limited to detecting apparent acceleration of a component in the radial direction.
In order to check for possible nonradial (non-ballistic) trajectories of components, we have also fit first-order 
linear functions separately to $x(t)$ and $y(t)$, as was also done for the 2~cm survey sources by K04.
These fits then determine a direction of motion, as well as the magnitude of the vector velocity.
They allow determination of possible nonradial trajectories by comparison of the fitted direction of motion with the
average position angle of the component.
Such non-ballistic trajectories may be caused by flow of the jet plasma along a curving channel, and such trajectories
have previously been detected in numerous individual source studies 
(dating back to at least Zensus, Cohen, \& Unwin [1995], for quasar 3C~345).
For consistency, we use here the same criteria as K04 
for determining what constitutes a significant detection of non-radial motion; namely, that the fit must meet the following
conditions: the component must be detected at at least five epochs, the vector velocity magnitude must be of at least 5$\sigma$
significance, and the fitted direction of motion must differ from the mean position angle of the component by at least 3$\sigma$.
A total of eighteen components satisfy the first two criteria, and of these six have significant nonradial motion
according to the third criterion. 
A third of the best-determined motions in the current RRFID sample are nonradial, the same fraction that was found for the 
2~cm survey by K04. The fit values for these six components are tabulated in Table~\ref{nonradialtab}.
Some components that are apparently stationary when only radial motion is considered may turn out to have
significant velocity magnitudes when the total velocity is taken into account (e.g., C2 in 0823+033 and C1 in 1611+343).
Of the six sources listed in Table~\ref{nonradialtab}, five are also in the 2~cm survey
(all except for 1622-253), but for 1611+343 and 2136+141 the
component listed in Table~\ref{nonradialtab} lies farther out than the components tracked in those sources in the 2~cm survey (K04).
For 0234+285, K04 also detect nonradial motion in the same component, and the parameters of the nonradial fits
are in good agreement. For 2200+420, K04 detect nonradial motion in the same direction as that listed in
Table~\ref{nonradialtab}, but in a different component. For 0823+033, our identification of components
differs from that by K04 (as discussed further in $\S$\ref{comp}), so that there is no corresponding detection of
nonradial motion in this source by K04.
Regardless of the agreement or lack thereof on specific sources, these two large VLBI surveys agree that roughly one third
of well-measured component trajectories are nonradial,
conclusively ruling out a purely ballistic model of the motion of radio-emitting components.

\begin{table*}[!ht]
\caption{Nonradial Trajectories}
\vspace{-0.20in}
\label{nonradialtab}
\begin{center}
\begin{tabular}{l c r r r r} \colrule \colrule
& & \multicolumn{1}{c}{$\overline{\mbox{PA}}$\tablenotemark{a}} & \multicolumn{1}{c}{$\phi$\tablenotemark{b}} & 
& \multicolumn{1}{c}{$|\overline{\mbox{PA}}-\phi|$} \\
\multicolumn{1}{c}{Source} & Comp. &
\multicolumn{1}{c}{(deg)} & \multicolumn{1}{c}{(deg)} & \multicolumn{1}{c}{$\beta_{app}$\tablenotemark{c}} &
\multicolumn{1}{c}{(deg)} \\ \colrule
0234+285 & C2 & $-13.5\pm0.3$  & $3.2\pm5.0$    & $10.2\pm1.4$ & $16.6\pm5.0$  \\
0823+033 & C2 & $26.4\pm1.3$   & $100.6\pm20.9$ & $2.4\pm0.4$  & $74.2\pm21.0$ \\
1611+343 & C1 & $163.5\pm0.2$  & $-117.2\pm7.1$ & $11.2\pm1.1$ & $79.3\pm7.1$  \\
1622-253 & C1 & $-11.0\pm1.5$  & $-66.9\pm16.1$ & $20.6\pm3.2$ & $55.9\pm16.1$ \\
2136+141 & C3 & $-104.3\pm2.6$ & $-151.3\pm5.9$ & $45.7\pm6.5$ & $47.0\pm6.5$  \\
2200+420 & C3 & $-167.9\pm1.8$ & $161.5\pm9.1$  & $3.0\pm0.6$  & $30.6\pm9.3$  \\ \colrule   
\end{tabular}
\end{center}
$a$: Average position angle of the component.\\
$b$: Fitted direction of motion of the component.\\
$c$: Magnitude of the velocity vector of the component.
\end{table*}

\subsection{Gamma-Ray Sources}
\label{gamma}
The detection by the EGRET instrument of of order 100
blazars at GeV gamma-ray energies (Hartman et al. 1999) opened up a new wavelength region to blazar astrophysics.
Interestingly, many very bright radio blazars were {\em not} detected above the EGRET threshold in GeV gamma-rays
(including, for example, 4C~39.25 from this survey),
raising the question of why many radio-loud blazars had strong, detectable gamma-ray emission
while others did not. One possible explanation for this is that the gamma-ray and radio
emission are relativistically beamed by different powers of the Doppler factor, with the gamma-ray emission being the more highly beamed.
This enhancement of Doppler beaming in gamma-rays
occurs in the synchrotron self-Compton (SSC) model for gamma-ray emission simply because of the steeper spectral index
in the gamma-ray portion of the spectrum. In the external-radiation Compton (ERC) model for gamma-ray emission, the beaming
enhancement of the gamma-ray emission relative to the radio emission can be even larger (Dermer 1995).
If the gamma-ray emission is more highly beamed than the radio emission,
then the EGRET blazars should be biased   
toward higher Doppler factors than radio-selected samples.
Assessing the effect of differing Doppler factors on observed apparent speeds is not trivial ---
interior to the critical angle that maximizes apparent speed the Doppler factor will be negatively correlated
with apparent speed, while outside this angle the correlation will be positive; so that one cannot simply conclude
that higher Doppler factors will produce faster jets.
Instead, these effects must be studied through Monte-Carlo simulations of the populations under consideration.
Simulations by Lister (1999), assuming a linear
relation between radio and gamma-ray luminosity, have shown that, in a radio-selected flux-limited sample, the
EGRET detected subset should indeed have systematically faster apparent speeds than the non-detections.

A study of the kinematic properties of the EGRET blazar jets was performed by Jorstad et al. (2001), using multi-epoch
VLBA images at 22 and 43 GHz. By comparing the apparent speeds measured in their study with those measured in the
radio-selected CJF sample, Jorstad et al. (2001) concluded that the EGRET sources were significantly faster 
than the radio-selected sources, and therefore more highly beamed. We have confirmed such an apparent speed difference in this
paper --- as discussed in $\S$\ref{appspeed}, a K-S test shows that the RRFID apparent speed distribution in Figure~4 differs
from the apparent speed distribution for the EGRET blazars in Jorstad et al. (2001) with high significance. However, the
measurements by Jorstad et al. (2001) were made at the high linear resolutions afforded by their 22 and 43 GHz observations,
and they therefore sampled regions closer to the core than either the CJF survey or the RRFID survey
in this paper. Separating possible resolution effects
from population effects is problematic, and ideally the gamma-ray and radio-selected samples should be observed at the
same frequencies. K04 split the 2~cm survey sources into two groups based on their EGRET detection
or non-detection, and found that the apparent speed distributions of the two groups differed, but at a relatively
low significance level of 90\%. Here we perform a similar calculation for the sources in the RRFID kinematic survey.
Of the 54 sources represented in the fastest-component histogram in Figure~5, 16 are EGRET detections according to
the analyses by Mattox, Hartman, \& Reimer (2001) and Sowards-Emmerd, Romani, \& Michelson (2003).
Figure~7 shows the histogram in Figure~5 separated according to EGRET detection, with the top panel representing
the EGRET detections and the bottom panel the non-detections. 
There is no significant statistical difference between the means of the two distributions shown in Figure~7.
A K-S test also shows no significant difference between the fastest apparent speeds of the EGRET and non-EGRET sources.
However, neither the 2~cm survey nor the RRFID survey represent a complete sample of gamma-ray sources, so both will be biased toward
those EGRET sources with higher radio fluxes. The issue is likely to remain open 
until large flux-limited samples in both wavebands are studied with VLBI with identical experimental setups.
Such studies will be significantly improved by the coming launch of GLAST, which should greatly increase the size
of gamma-ray selected blazar samples.

\begin{figure*}[!ht]
\begin{center}
\includegraphics[scale=0.50]{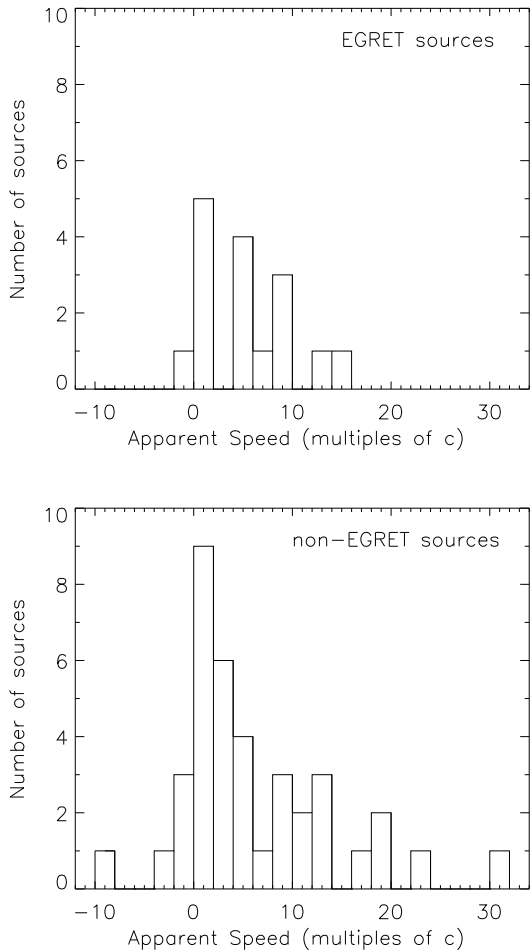}
\end{center}
\vspace{-0.20in}
\caption{Histograms of the fastest apparent component speed in each source,
for the 54 sources represented in Figure~5, separated by EGRET detection status. 
EGRET sources from Mattox et al. (2001) and Sowards-Emmerd et al. (2003) are in the upper panel (16 sources), sources not detected
by EGRET are in the bottom panel (38 sources).}
\end{figure*}


\section{Source-by-Source Comparison to 2~cm Survey Results}
\label{comp}
As discussed in $\S$\ref{data}, the 2~cm survey observations described by K04 and the RRFID
kinematic survey observations described here are nearly matched in angular resolution and sensitivity.
Of the 77 sources included in Table~4, 36, or about half, also have measured apparent speeds published by K04 from the 2~cm survey data.
This provides a unique opportunity to compare the results from the two surveys for the common sources, in order to
see if the standard procedure of VLBI model-fitting, component identification, and linear fitting
yields reproducible results for the apparent speeds when the same sources are analyzed using different datasets by different groups.
To aid in this comparison, 
all kinematic results in this paper were produced `blind' with respect to the 2~cm survey results, and the
comparisons in this section were not made until after the kinematic results in $\S$\ref{appspeed}
had been finalized.
We note that for some of the sources there are also numerous other published multi-epoch VLBI results, 
but since those are typically over different
time ranges or at different resolutions, we restrict the source-by-source comparison in this section
to a comparison between results from this paper and those by K04.

Figure~8 shows the source-by-source comparison of the model-fit component positions, component identifications,
and fitted apparent speeds for the 36 common sources in the RRFID kinematic survey and the 2~cm survey.
The data and fits plotted in black in Figure~8 are the RRFID results transposed from Figure~2. The data and fits plotted
in red in Figure~8 are the 2~cm survey results from Figure~1 of K04.
Note first that there are many very nice cases of agreement between the two surveys --- we point out the results for 1128+385 and 1308+326
as specific examples where the independent models from the two surveys are so close to each other 
on Figure~8 as to be nearly indistinguishable.
However, there are other cases where the component identifications and measured apparent speeds are quite different --- this issue
is discussed in more detail below.
There is no systematic detection of a consistent frequency-dependent separation of components from the cores in Figure~8; such
frequency-dependent separation would be expected if sources have optically thick surfaces (`cores') at 8 and 15 GHz that are a significant 
distance apart. In most cases where the identification schemes agree, the separations 
at the two frequencies are the same within the errors, as in the 
two sources mentioned specifically above. We do note two cases of apparent frequency-dependent separation: in 2234+282 where the single
component is about 0.5 mas farther from the `core' at 15 GHz, and in 0119+041 where the single component is about 0.2 mas closer to the
`core' at 15 GHz (opposite the usual expected sense for frequency-dependent separation).
In any event, any such frequency-dependent separations would not effect the apparent speed measurements, but only the estimated
epoch of ejection from the core.

\begin{figure*}
\begin{center}
\includegraphics[scale=0.975,angle=180.0]{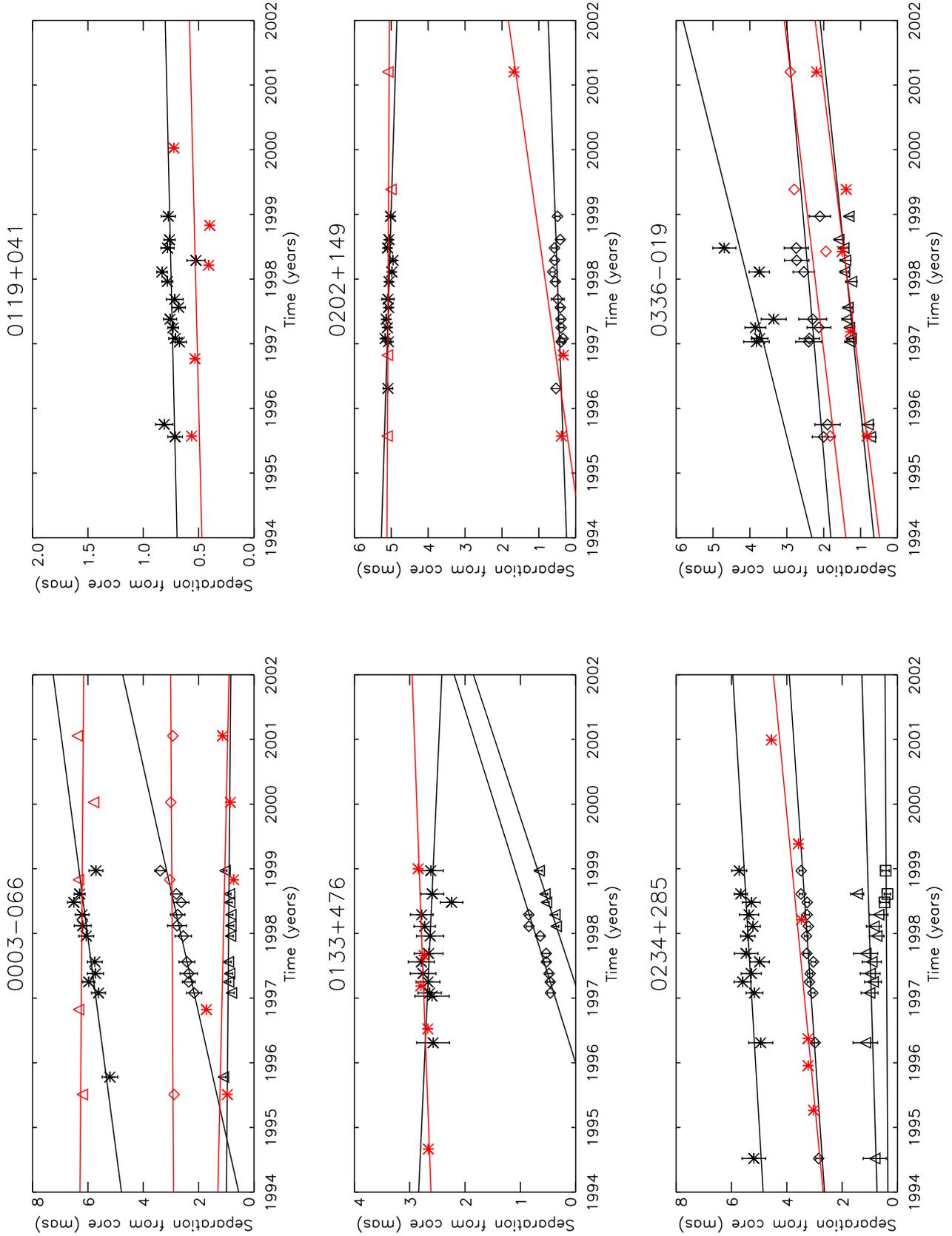}
\end{center}
\vspace{-0.40in}
\caption{Comparison of model-fit component positions and fitted apparent speeds for the common sources
in the RRFID kinematic survey and 2~cm surveys.
The data and fits plotted in black are the results from this paper transposed from Figure~2. The data and fits plotted
in red are the 2~cm survey results from Figure~1 of K04.
For the RRFID data, component symbols are the same as those used in Figure~2.
For the 2~cm survey data, asterisks are used to represent component B, diamonds component C, triangles component D,
squares component E, and x's component F, as identified by K04.}
\end{figure*}

\begin{figure*}
\begin{center}
\includegraphics[scale=0.975,angle=180.0]{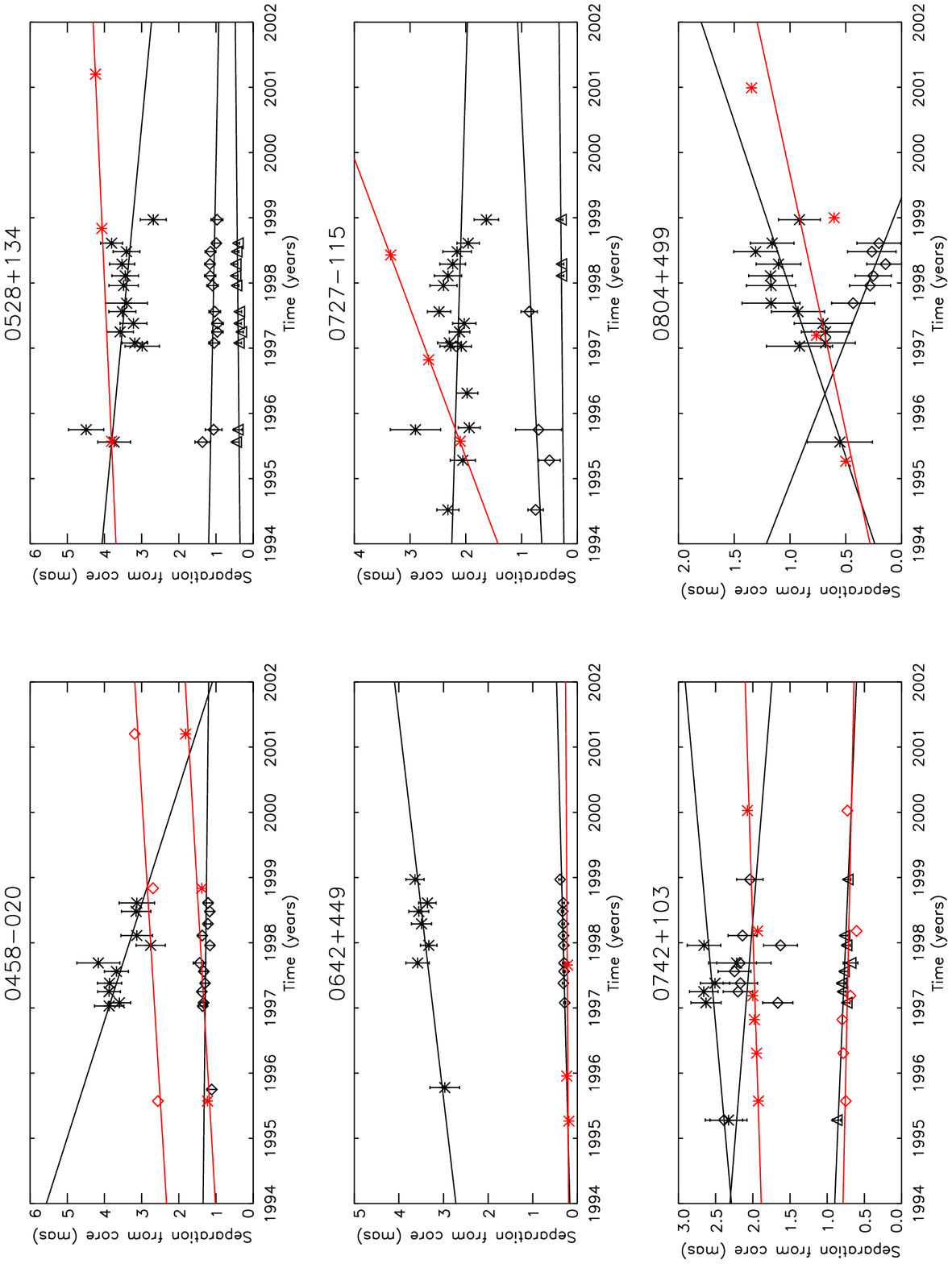}
FIG. 8.---{\em Continued}
\end{center}
\end{figure*}

\begin{figure*}
\begin{center}
\includegraphics[scale=0.975,angle=180.0]{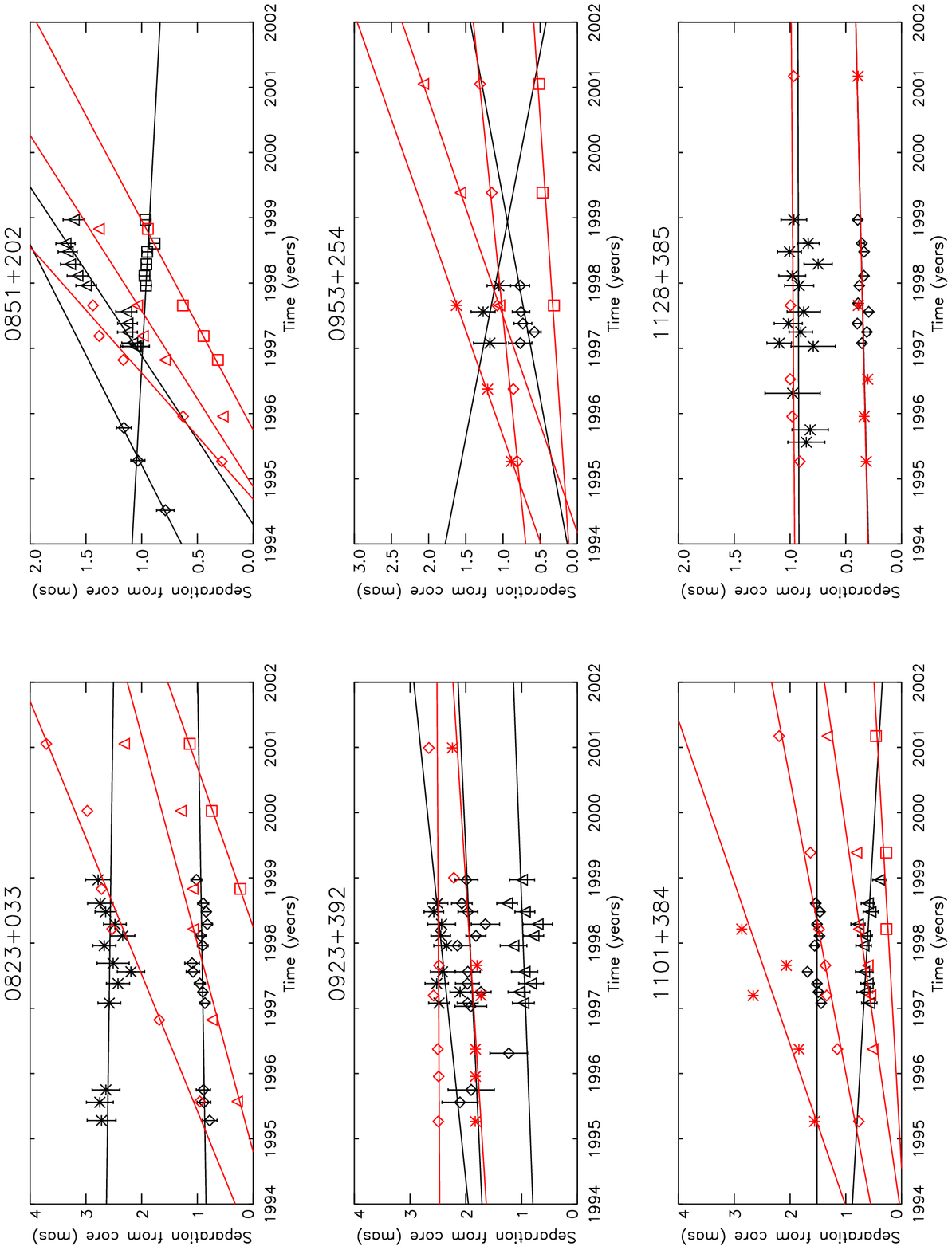}
FIG. 8.---{\em Continued}
\end{center}
\end{figure*}

\begin{figure*}
\begin{center}
\includegraphics[scale=0.975,angle=180.0]{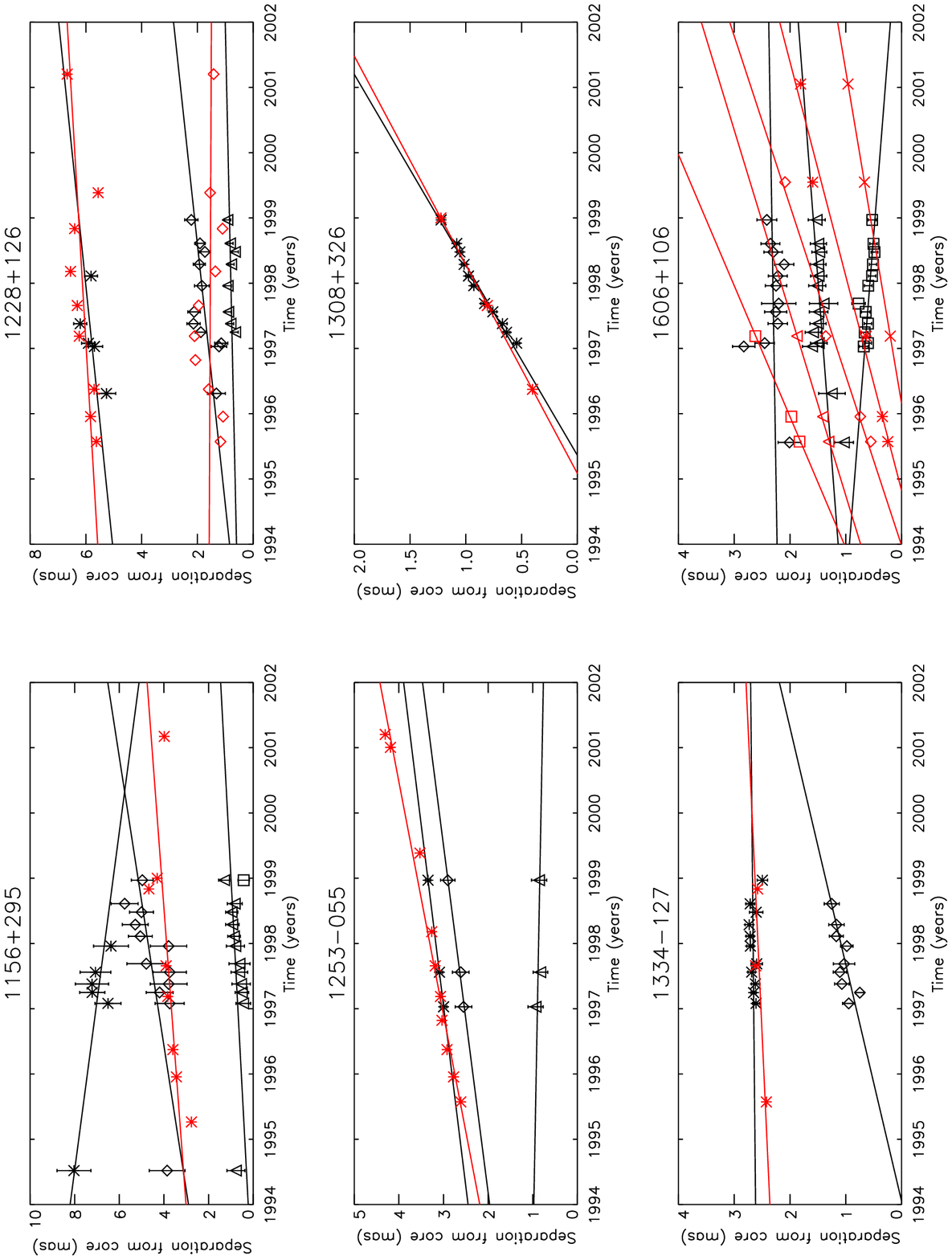}
FIG. 8.---{\em Continued}
\end{center}
\end{figure*}

\begin{figure*}
\begin{center}
\includegraphics[scale=0.975,angle=180.0]{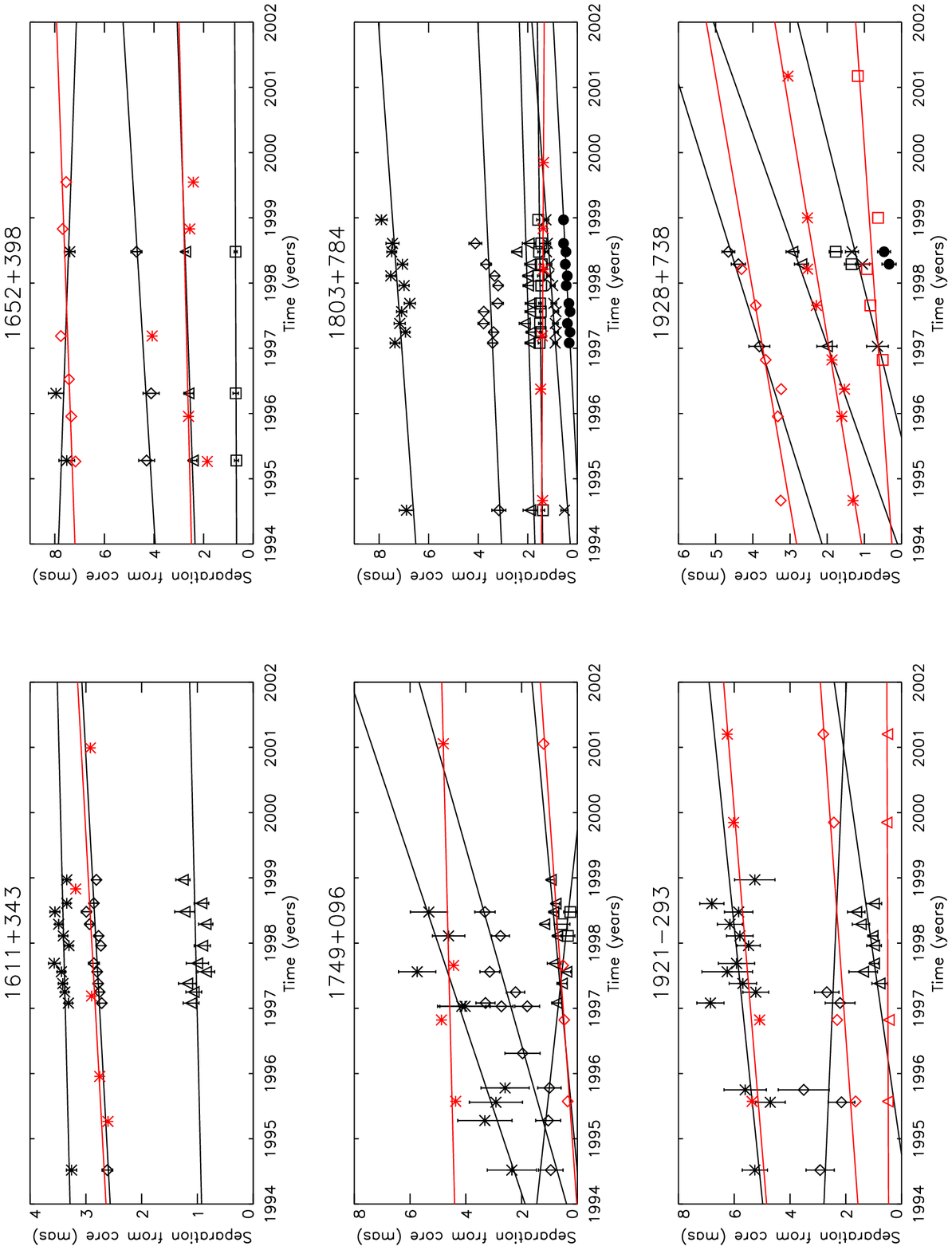}
FIG. 8.---{\em Continued}
\end{center}
\end{figure*}

\begin{figure*}
\begin{center}
\includegraphics[scale=0.975,angle=180.0]{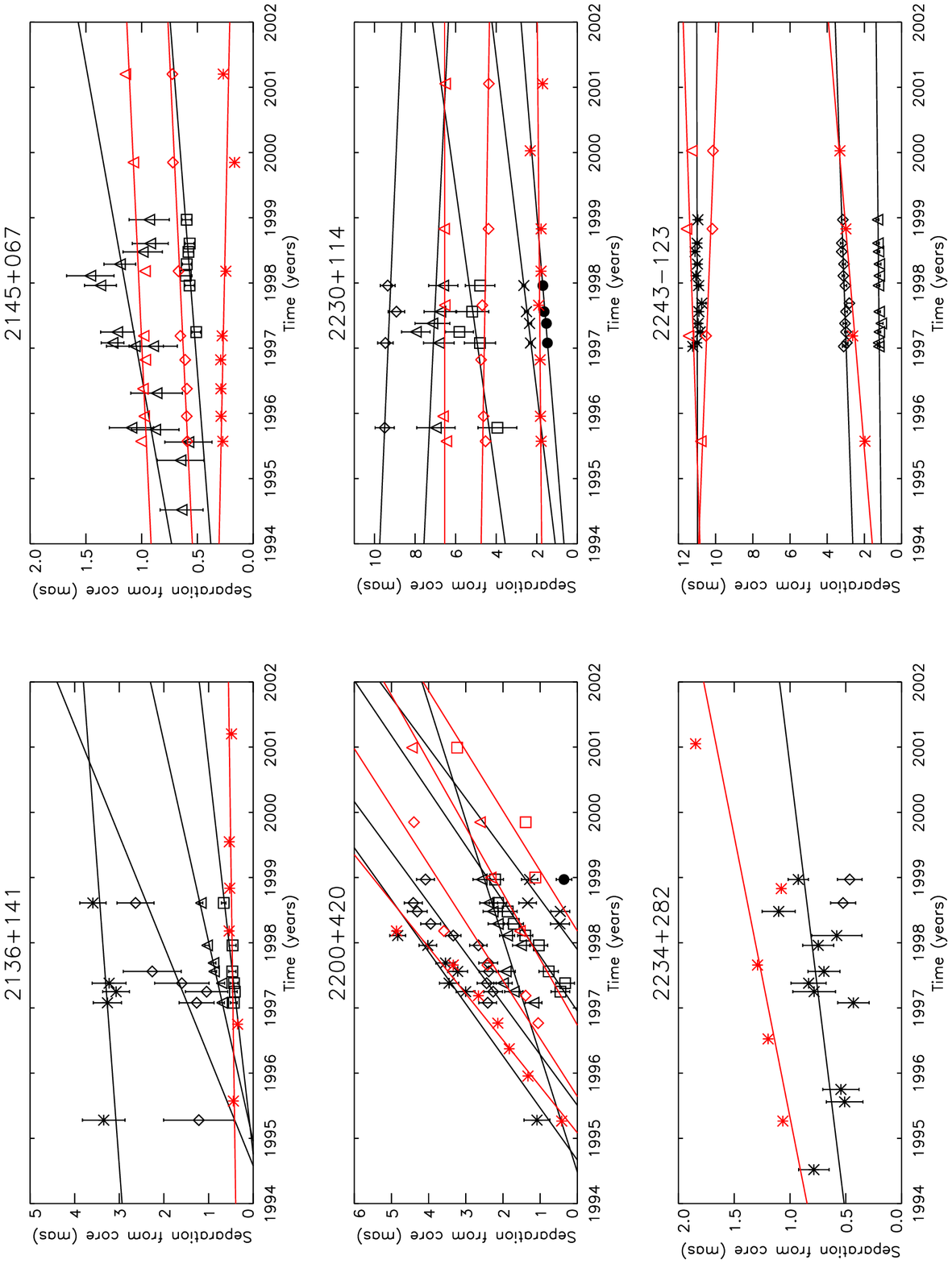}
FIG. 8.---{\em Continued}
\end{center}
\end{figure*}

In these 36 sources, there are 66 components from the RRFID kinematic survey that can, at at least one epoch, be matched   
with a corresponding component from the same source from K04. 
From these 66 components, we list the 40 components (from 28 sources) with `Good' or `Excellent' quality codes in {\em both}
surveys, together with their measured proper motions,
in Table~\ref{comptab} (comparing proper motions rather than speeds avoids the issue
of the slightly different cosmologies assumed by the two surveys).
In a number of sources, different component identification schemes were used for a given source in the two surveys, making the identification
of which component from Table~4 `matches' which component from K04 problematic (the problem being that
what is identified as a single component in one survey may have been identified as different components at different times in
the other survey). In these cases, the correspondence was made using the best-matching component pairs from the two datasets
during the time period 1997-1998, when the RRFID observations are the most densely sampled.

\begin{table*}
\label{comptab}
\caption{Comparison of Measured Proper Motions}
\vspace{-0.20in}
\begin{center}
\begin{tabular}{l c r c r c c} \colrule \colrule
& & \multicolumn{1}{c}{RRFID\tablenotemark{a}} & & \multicolumn{1}{c}{2 cm\tablenotemark{b}} \\
& RRFID\tablenotemark{a} & \multicolumn{1}{c}{Proper Motion} & 2 cm\tablenotemark{b} &
\multicolumn{1}{c}{Proper Motion} & Difference\tablenotemark{c} & Cause of\tablenotemark{d} \\
\multicolumn{1}{c}{Source} & Component & \multicolumn{1}{c}{(mas/yr)} &
Component & \multicolumn{1}{c}{(mas/yr)} & in $\sigma$  & Disagreement \\ \colrule
0003-066 & 1 & $0.308\pm0.091$  & D & $-0.02\pm0.06$  & {\bf 3.0}  & MF,TB \\ 
         & 2 & $0.524\pm0.124$  & C & $0.01\pm0.02$   & {\bf 4.1}  & MF,TB \\ 
         & 3 & $-0.020\pm0.029$ & B & $-0.05\pm0.09$  & 0.3  \\ 
0119+041 & 1 & $0.013\pm0.018$  & B & $0.01\pm0.04$   & 0.1  \\ 
0133+476 & 1 & $-0.052\pm0.093$ & B & $0.04\pm0.01$   & 1.0  \\ 
0202+149 & 1 & $-0.053\pm0.034$ & D & $-0.01\pm0.01$  & 1.2  \\
0234+285 & 2 & $0.159\pm0.023$  & B & $0.23\pm0.05$   & 1.3  \\
0336-019 & 3 & $0.183\pm0.037$  & B & $0.22\pm0.04$   & 0.7  \\
0458-020 & 2 & $-0.019\pm0.044$ & B & $0.10\pm0.04$   & 2.0  \\
0528+134 & 1 & $-0.166\pm0.111$ & B & $0.077\pm0.002$ & 2.2  \\
0727-115 & 1 & $-0.035\pm0.044$ & B & $0.44\pm0.01$   & {\bf 10.5} & MF \\
0742+103 & 2 & $-0.071\pm0.078$ & B & $0.03\pm0.01$   & 1.3  \\
         & 3 & $-0.037\pm0.018$ & C & $-0.02\pm0.02$  & 0.6  \\
0804+499 & 1 & $0.195\pm0.080$  & B & $0.13\pm0.06$   & 0.7  \\
0823+033 & 1 & $-0.016\pm0.056$ & C & $0.48\pm0.04$   & {\bf 7.2}  & MF,ID \\
         & 2 & $0.019\pm0.027$  & D & $0.31\pm0.06$   & {\bf 4.4}  & ID \\
0851+202 & 3 & $0.386\pm0.043$  & D & $0.37\pm0.06$   & 0.2  \\
         & 4 & $-0.031\pm0.059$ & E & $0.31\pm0.02$   & {\bf 5.5}  & ID \\
0923+392 & 2 & $0.053\pm0.070$  & B & $0.07\pm0.03$   & 0.2  \\
1101+384 & 2 & $0.000\pm0.041$  & C & $0.22\pm0.02$   & {\bf 4.8}  & ID \\
         & 3 & $-0.067\pm0.068$ & D & $0.17\pm0.03$   & {\bf 3.2}  & ID \\
1128+385 & 1 & $0.001\pm0.038$  & C & $0.004\pm0.008$ & 0.1  \\
         & 2 & $0.014\pm0.016$  & B & $0.01\pm0.01$   & 0.2  \\
1253-055 & 1 & $0.181\pm0.046$  & B & $0.28\pm0.01$   & 2.1  \\
1308+326 & 1 & $0.343\pm0.014$  & B & $0.313\pm0.002$ & 2.1  \\ 
1606+106 & 3 & $0.089\pm0.049$  & C & $0.38\pm0.03$   & {\bf 5.1}  & ID \\ 
         & 4 & $-0.092\pm0.032$ & B & $0.30\pm0.02$   & {\bf 10.4} & ID \\ 
1611+343 & 2 & $0.063\pm0.022$  & B & $0.06\pm0.04$   & 0.1  \\ 
1803+784 & 4 & $0.027\pm0.017$  & B & $-0.01\pm0.01$  & 1.9  \\ 
1921-293 & 1 & $0.241\pm0.110$  & B & $0.19\pm0.06$   & 0.4  \\ 
2136+141 & 4 & $0.167\pm0.044$  & B & $0.02\pm0.01$   & {\bf 3.3}  & ID \\ 
2145+067 & 3 & $0.105\pm0.036$  & D & $0.03\pm0.01$   & 2.0  \\ 
	 & 4 & $0.046\pm0.030$  & C & $0.027\pm0.003$ & 0.6  \\ 
2200+420 & 1 & $1.253\pm0.136$  & B & $1.41\pm0.13$   & 0.8  \\ 
         & 2 & $1.288\pm0.128$  & C & $1.12\pm0.22$   & 0.7  \\ 
	 & 4 & $1.183\pm0.146$  & D & $0.99\pm0.18$   & 0.8  \\ 
2230+114 & 6 & $0.265\pm0.130$  & B & $0.03\pm0.04$   & 1.7  \\ 
2234+282 & 1 & $0.072\pm0.031$  & B & $0.12\pm0.05$   & 0.8  \\ 
2243-123 & 1 & $0.006\pm0.060$  & D & $0.11\pm0.10$   & 0.9  \\ 
         & 2 & $0.118\pm0.033$  & B & $0.29\pm0.03$   & {\bf 3.9}  & TB \\ \colrule
\end{tabular}
\end{center}
$a$: Component identifications and proper motions from Table~4 of this paper.\\
$b$: Component identifications and proper motions from Table~2 of K04.\\
$c$: The significance of the difference in the two proper motions was computed from the
probability associated with the reduced chi-squared that was computed from the 
fit of the two proper motions to their weighted average.
Entries in bold indicate a probability $p<0.01$ that the two proper motions are independent measurements of the same proper motion.\\
$d$: For the bold entries in the previous column, this column indicates the cause or causes of the different
proper motion measurements in the two surveys: MF -- the model fits give different measured component positions in the two
surveys at similar epochs, TB -- measurements on different time baselines have caused different fitted proper motions, ID --
different component identification schemes were used by the two surveys. 
See the text in $\S$~\ref{comp} for a discussion of each of these.
\end{table*}

The measured proper motions from this paper are plotted versus the corresponding proper motions from the 2~cm survey in Figure~9,
using the 40 components from Table~\ref{comptab}.
The proper motions from the two surveys show a correlation with high significance (linear correlation coefficient of 0.81);
however, this correlation is mainly due to the similar measurements for the proper motions of three components in BL Lac, which lie
in the upper right corner corner of the figure. If those three components are excluded, then there is no statistically significant
correlation between the proper motions measured by the two surveys. 
This lack of correlation is caused by the presence of a non-negligible subset of components
with systematic differences between their proper motion measurements in the two surveys --- this subset
of components is identified and discussed in detail below.
If that subset of 12 components is removed from Figure~9, then a correlation with high significance 
is recovered, even when the three components of BL Lac are excluded (linear correlation coefficient of 0.76). 

\begin{figure*}
\begin{center}
\includegraphics[scale=0.60]{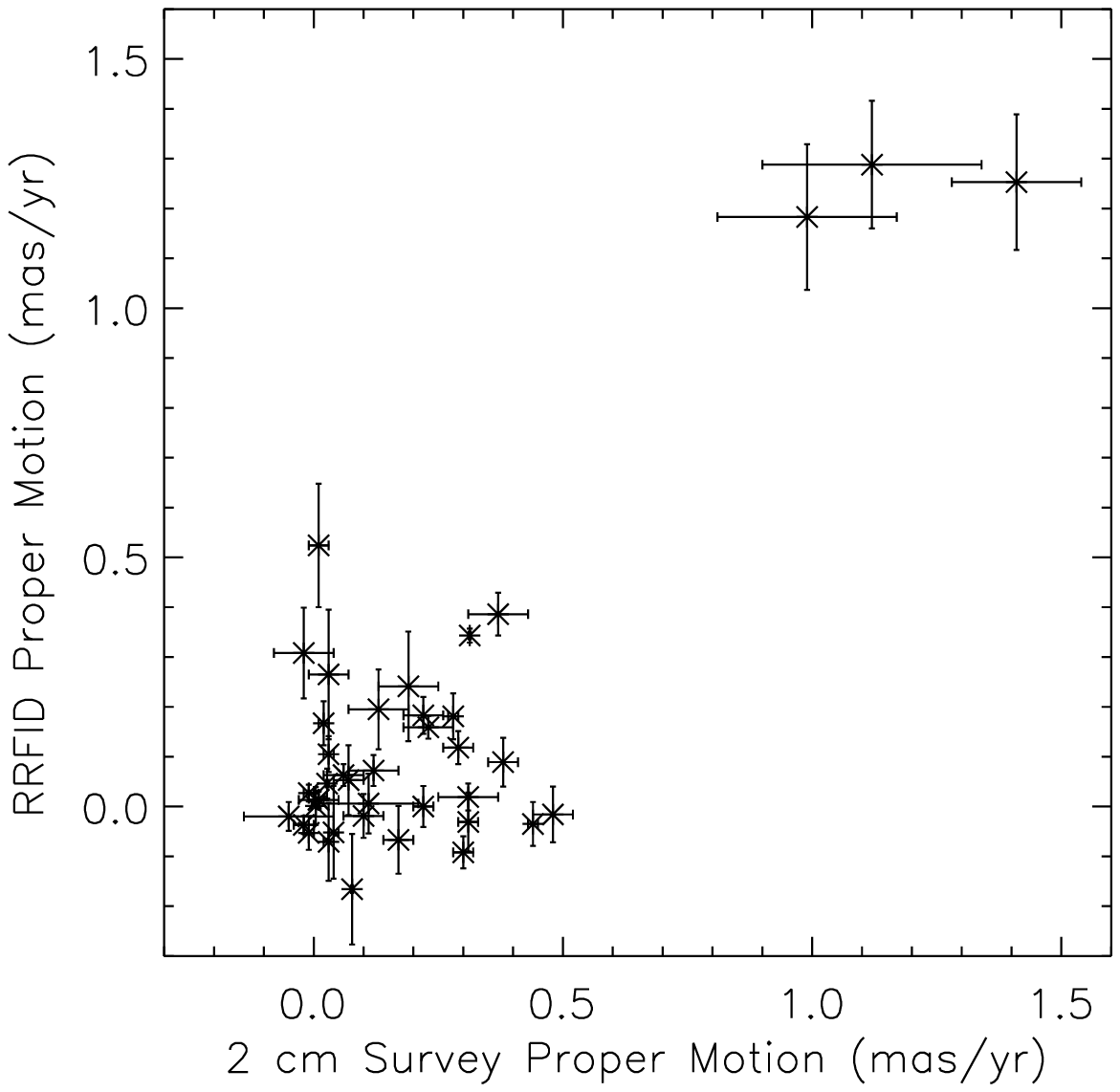}
\end{center}
\caption{Measured proper motion from the RRFID kinematic survey (this paper) versus the measured proper motion
from the 2~cm survey (K04), for the common components listed in Table~\ref{comptab}.
Some error bars are smaller than the plotting symbols.}
\end{figure*}

We evaluated the agreement between the two surveys on a component-by-component basis by computing the significance
of the difference in the two proper motion measurements for the `matching' components, this significance is tabulated
for each component in Table~\ref{comptab}. The significance of the difference in the two proper motions was computed from the 
probability associated with the reduced chi-squared, computed from the fit of the two proper motions to their weighted average.
For 12 of the 40 entries in Table~\ref{comptab} (or about 25\% of the entries), 
the calculated significance exceeds 2.6$\sigma$, corresponding to
a probability $p<0.01$ that the two proper motions are independent measurements of the same proper motion
(these significances are listed in bold type in Table~\ref{comptab}).
For these 12 components at least, there are some systematic errors in the interpretation of the VLBI data that
are having an effect. Differences for the remaining 28 components follow approximately the expected normal distribution,
as shown in Figure~10.
In the rest of this section, we attempt to identify the types of systematic errors that have affected the 12 outlying components.
We have classified these systematic errors into three types, and a code is given for the type of error or errors applicable to each 
of these 12 components in the last column of Table~\ref{comptab}. Below we discuss these types of errors in detail, with specific
examples of each:

\begin{figure*}
\begin{center}
\includegraphics[scale=0.60]{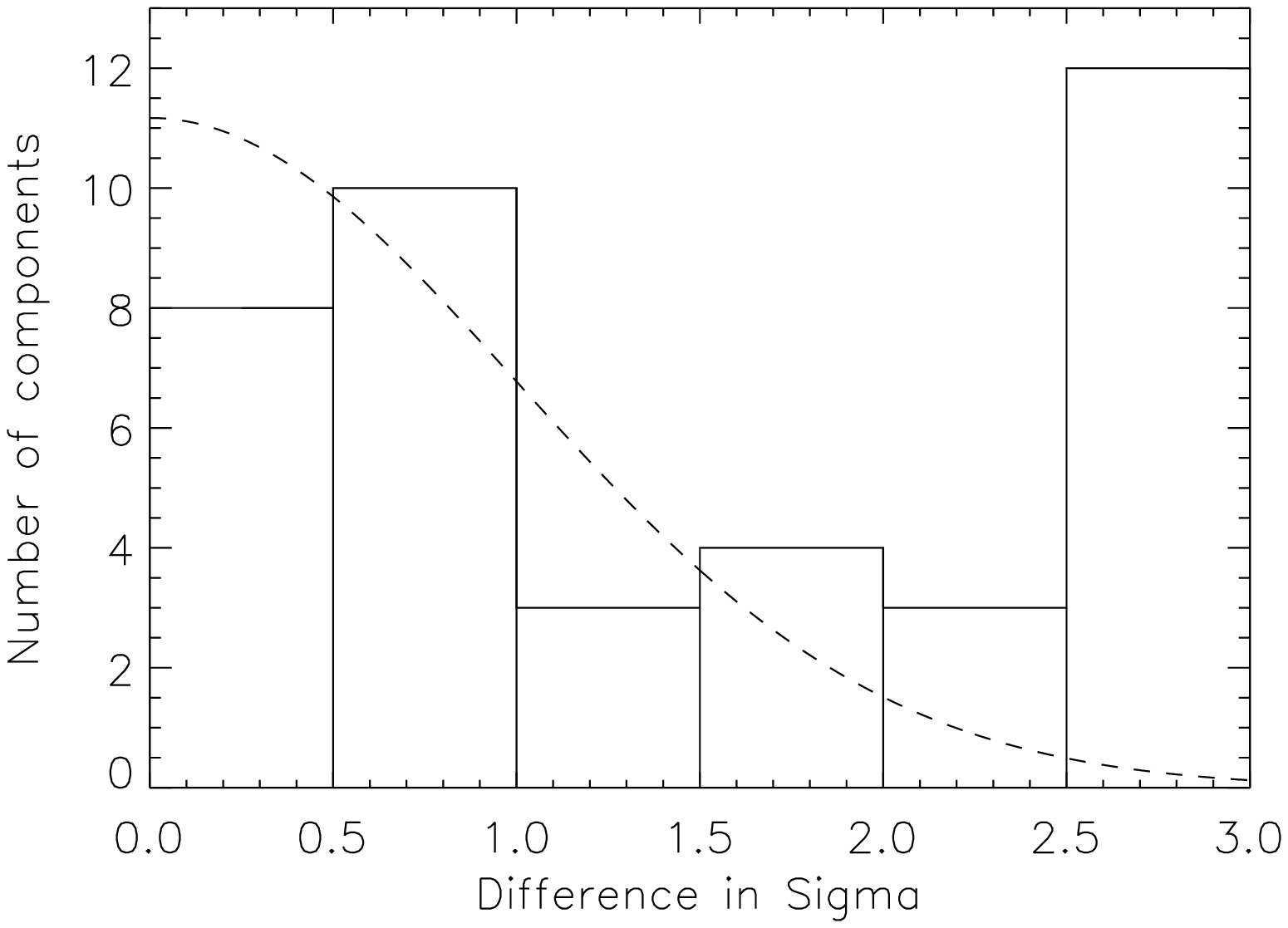}
\end{center}
\caption{Histogram of the difference in sigma between the RRFID proper motion measurement and the 2~cm survey
proper motion measurement for the 40 common components in Table~\ref{comptab}.  The 12 components with
$>2.6\sigma$ difference are all included in the rightmost bin. The dashed line shows the theoretical normal
distribution for the remaining 28 components.}
\end{figure*}

{\bf Model fit differences:} In the majority of cases where the RRFID kinematic survey and the 2~cm survey have data on
the same source at nearly the same epoch, then the fitted positions of the model components agree between the two surveys.
However, in a small set of cases there is significant disagreement. For example, in the 1995 epochs for 0003-066 in Figure~8,
K04 measure the outermost component to be about 1 mas farther from the core, and they measure an extra component in between the two
components measured by this paper. (At later epochs for this same source, e.g., 1998, the two surveys get nearly identical results.)
Similar cases are indicated by the `MF' code in the final column of Table~\ref{comptab}.
Interestingly, the three sources which are noted as having different fitted component positions between the two surveys
(0003-066, 0727-115, and 0823+033) are also among the 10\% of sources mentioned in 
$\S$~\ref{modelfitting} for which we obtain significantly different component positions
from those obtained from the RRFID data by Fey et al. (1996), Fey \& Charlot (1997), and Fey \& Charlot (2000).
If the corresponding fits from those references are added to Figure~8 for these
sources, then they represent a third possible deconvolution of the source structure that does not agree with the other two.
In these cases, the source structure is apparently ambiguous, and
the deconvolution obtained is apparently quite sensitive to the dataset and to the details of the modelfitting approach used.

{\bf Time baseline differences:} In some sources, the RRFID kinematic survey and the 2~cm survey obtain similar model-fit
positions at epochs when both surveys have data, but positions measured at other times when only one of the two surveys has data
cause differences in the apparent speed measurements. 
The second component in 2243-123 is an example of this. 
The first measurement of the position of this component by K04 in 1995 lies considerably below the extrapolation of 
the RRFID fit, as if the component decelerated sometime during the years 1995 to 1997.
In fact, if components do not accelerate or decelerate, then differences in the time baseline should have no
affect on the apparent speed measurements (other than to reduce the random error), so we conclude that such components
are most likely cases of apparent acceleration or deceleration of the radial motion.
Such cases are indicated by the `TB' code in the final column of Table~\ref{comptab}.
There are additional examples of this, such as the inner component in 0202+149, that did not meet the criteria
for inclusion in Table~\ref{comptab}.

{\bf Different component identification schemes:} A subset of sources exists where both the 2~cm survey and the RRFID kinematic survey
have measured similar positions for model-fit components at similar epochs, 
but where the components have been identified differently from epoch to epoch to
yield different sets of `components' with different speeds. The source 1606+106 in Figure~8 provides an example of this.
If one considers the data for component 3 
in this source from the RRFID kinematic survey (black triangles on Figure~8 and associated black-line fit),
then it can be seen that there is a matching 2~cm survey component in 1996 (identified as their component D), another
in 1997 (identified as their component C), and another in 1999 (identified as their component B). 
This is typical of this general problem --- one survey has interpreted the position data as a smaller number of more slowly moving components,
while the other has interpreted it with a larger number of faster components.
In these cases, it is the spacing of the epochs in time that is the major influence on component identification.
For 1606+106, the dense spacing of the RRFID position data during 1997 and 1998 seems to preclude the faster interpretation of K04.
Four other sources where differing component identification schemes have yielded significantly different apparent speed measurements
are indicated by the `ID' code in the final column of Table~\ref{comptab}. 
None of these components were recorded as `Distinct Features' in {\em both} surveys, so that seeing
a component as a distinct rather than a blended feature on the images evidently helps prevent ambiguities
in component identification. In addition, there are five other sources where components were identified differently by the two surveys,
but where the motions did not meet the criteria for inclusion in Table~\ref{comptab}, or where 
this did not result in apparent speed differences at the cutoff significance level specified for the analysis in this section.
In total, different component identification schemes were used for 10 of the 36 common sources.
For those discrepant sources that have densely spaced RRFID observations during 1997 and 1998, 
we expect that the more densely spaced RRFID data should better resolve potential ambiguities in component identification.

Ambiguities in identifying VLBI components from epoch to epoch has long been discussed as a potential problem
for multi-epoch VLBI observations (see, for example, the discussion of this issue in the context of the 
CJF survey by Vermeulen et al. [2003]). However, this is the first time that this issue has been quantitatively addressed using 
independently analyzed datasets for a sizable number of common sources.
If the results from this paper can be extrapolated to similar multi-epoch VLBI studies,
then roughly 25\% of the apparent speed measurements in the literature may not be repeatable, in the sense that other observers
using a similar but independent set of VLBI observations may have reached different conclusions about the apparent speeds.
Component identification should be considered more reliable for distinct features on the images, and should also become more robust
as the time-density of epochs increases, because there are fewer consistent ways in which
the components can be identified. For multi-epoch VLBI surveys then, the distribution of apparent speeds
seems to be a robust and repeatable measurement (witness the statistical agreement between the apparent speed distributions
measured by the 2~cm survey and this paper, as discussed in $\S$\ref{appspeed}),
but we would caution against relying overly much on the apparent speed measured for a particular source, unless 
the component identification in that source is well constrained by having many more observed epochs than model components,
or by seeing that particular component as a strong, distinct feature on the images.

Whether or not two series of VLBI observations agree on their measurements of component motions is a separate
question from whether or not those motions are a realistic portrayal of what is going on in the jet.
Comparison of VLBA observations with simulations of relativistic jets by Gomez (2005)
led to the conclusion that the interpretation of VLBI images as a series of Gaussian moving components is an overly simplistic
idealization of more intricate jet emission patterns.
Some of the disagreements on `component' motions discussed above then probably arise from different approximations
to an underlying complex flow that cannot be fully resolved. However, because the VLBI data are only partially resolved,
models consisting of a series of Gaussians with a few free parameters
do fit the observed visibilities with reasonable reduced chi-squareds, so unless the linear
resolution of jet observations increases, it will be difficult to constrain fits to the more complex emission patterns
suggested by the numerical simulations. 
Despite these problems, the moving Gaussian approximation does provide some valuable information about the sources.
As shown here, the apparent speed measurements in this approximation are repeatable for about 75\% of the sources, and correlations between
the measured apparent speeds and other source properties (in, for example, the MOJAVE survey [Lister 2006])
have shown that the fastest measured apparent speeds in the
Gaussian approximation are a good realization of the bulk apparent speeds of the jets.

\section{Conclusions}
Some of the major conclusions from the present work are as follows:
\begin{description}
\item[1.]{The Radio Reference Frame Image Database has been validated as valuable tool 
for studying jet kinematics. All 8 GHz VLBA images in the RRFID for all
87 sources observed at 3 or more epochs over the years 1994-1998 were considered, and
in total we identified and measured apparent speeds for 
184 jet components in 77 of these sources, with an average of 11 epochs of observation per source.
About half of these sources are not present in other large multi-epoch VLBI surveys,
so that these results represent the first information on the jet kinematics in these sources.}
\item[2.]{The measured apparent speed distribution for the 94 best-measured components (Figure~4)
shows a peak at low apparent speeds that is consistent with a population of stationary components, a tail
extending out to apparent speeds of about $30c$, and a mean apparent speed of 3.6$c$.
The distribution is statistically consistent with the apparent speed distribution
found by the radio-selected 2~cm survey, but differs significantly from the
apparent speed distribution in gamma-ray blazars measured by Jorstad et al. (2001).}
\item[3.]{For the 36 sources in common between this survey and the 2~cm survey, we made a component-by-component
comparison of the measured apparent speeds. Significant disagreements are found in about 25\% of the apparent
speed measurements, usually due to different assumed component identification schemes. This first large-scale
test of the repeatability of apparent speed measurements shows that component identification can be a significant
problem that is probably best avoided by short spacings between observing epochs.}
\end{description}

Some other results are:
\begin{description}
\item[4.]{There is a difference between the fastest measured apparent
speeds in the quasars and BL Lac objects, with the quasar jets being faster, but the significance of this result
is a rather low 94\%.}
\item[5.]{We checked for accelerated radial motion and nonradial trajectories in the individual component position data.
We found significant nonradial motion for six components,
and report a tentative detection of accelerated radial motion for three components.}
\item[6.]{There was no significant statistical difference in the apparent speed distributions of the EGRET-detected
and non-detected sources;
however, this survey does not contain a complete gamma-ray or radio-selected sample.}
\end{description}

This paper has presented some of the first astrophysical results to be derived
from the Radio Reference Frame Image Database, and it is clear that this
database can provide scientific results that compare favorably with those of other large VLBI surveys. 
However, we reiterate that the RRFID in not a complete flux-limited sample, and that
the exact nature of the biases that this lack of pre-defined selection criteria may introduce into
statistical quantities calculated from the RRFID is not known.
This paper has only presented the apparent jet speed measurements made from the first five years of
the RRFID (1994-1998), with a minimal amount of further analysis.
Many more studies are possible with the current set of reduced RRFID data, and 
some of these will be pursued in future papers. Such studies include a paper in preparation on the
misalignment angles between the parsec and kiloparsec scale structures in the RRFID sources, 
and a future study of possible transverse jet structures. Those studies will take advantage of the 2 GHz data
present in the RRFID in addition to the 8 GHz data that was used in this paper, and will not be limited
to the sources observed at three or more epochs that we have restricted ourselves to here.

In addition, astrometric and geodetic VLBA observations (the RDV experiment series) have continued at the rate of six epochs per year
since the end of the RRFID data included in this paper in December 1998.
We are presently working as part of a collaborative effort to complete the imaging of the 30 RDV experiments
observed during the years 1999-2003, which will make the RRFID complete over a ten-year time
baseline, with approximately 50 epochs per source for the best-observed sources.
Once the RRFID has been updated with a longer time baseline,
we can update the apparent speed measurements given in this paper, significantly reducing the random errors.
Those updated apparent speeds can then be used for more detailed studies of jet physics, including
studies of correlations with other source properties, radio source evolution and unification, and cosmology.
With a long time baseline and dense epoch spacing providing up to 50 observed epochs per source, 
the continuation of the RRFID kinematic survey
will continue to provide a valuable comparison to other active VLBI surveys such as the MOJAVE and MOJAVE-II surveys.

\acknowledgments
We acknowledge insightful and relevant comments from the anonymous referee, which greatly
improved the paper.
We acknowledge the 2~cm survey team, in particular Ken Kellermann, for supporting
M. Mahmud during a summer internship at NRAO, and for kindly providing data in advance of publication.
The National Radio Astronomy Observatory is a facility of the National Science Foundation 
operated under cooperative agreement by Associated Universities, Inc.
Part of the work described in this paper has been carried out at the Jet
Propulsion Laboratory, California Institute of Technology, under
contract with the National Aeronautics and Space Administration.
This research has made use of the United States Naval
Observatory (USNO) Radio Reference Frame Image Database (RRFID), and the
NASA/IPAC Extragalactic Database (NED), which is operated by the Jet Propulsion Laboratory, 
California Institute of Technology, under contract with the National Aeronautics and Space Administration.
This work was supported by the
National Science Foundation under Grant No. 0305475,
and by a Cottrell College Science Award from Research Corporation.

\end{document}